\documentclass[aps,prx,reprint,superscriptaddress,longbibliography,nofootinbib]{revtex4-2}
\usepackage[colorlinks=true,hyperfootnotes=true,breaklinks=true,citecolor=red,urlcolor=blue,linkcolor=blue]{hyperref}
\usepackage{amsmath,amssymb,physics,enumerate,graphicx,placeins,bm,adjustbox,xcolor,amsthm,tabularx,dcolumn}

\begin{document}

\title{Current fluctuations in a non-additive open quantum system: breakdown of the quantum-jump approach}

\author{Ilia Khomchenko}
\email{ilia.khomchenko@um.edu.mt}
\affiliation{Department of Physics, University of Malta, Msida MSD 2080, Malta }
\author{Saulo~V.~Moreira}
\email{moreirsv@tcd.ie}
\affiliation{School of Physics, Trinity College Dublin, College Green, Dublin 2, D02 K8N4, Ireland}
\author{Emanuel Schwarzhans}
\email{emanuel.schwarzhans@um.edu.mt}
\affiliation{Department of Physics, University of Malta, Msida MSD 2080, Malta }
\affiliation{Atominstitut, TU Wien, 1020 Vienna, Austria}
\author{Mark~T.~Mitchison}
\email{mark.mitchison@kcl.ac.uk}
\affiliation{School of Physics, Trinity College Dublin, College Green, Dublin 2, D02 K8N4, Ireland}
\affiliation{Department of Physics, King’s College London, Strand, London, WC2R 2LS, United Kingdom}
\author{Tony~J.~G.~Apollaro}
\email{tony.apollaro@um.edu.mt}
\affiliation{Department of Physics, University of Malta, Msida MSD 2080, Malta }

\date{\today}
\begin{abstract}
Open quantum system dynamics is efficiently described by the quantum master equation formalism. Therein, quantum master equations in Lindblad form constitute an important subclass describing Markovian dynamics. When an open quantum system is in an out-of-equilibrium state, an exchange of particles between the open system and reservoirs takes place yielding to a non-zero average net current and associated current fluctuations, which can be characterised with the quantum jump formalism for quantum master equations expressed in Lindblad form. However, a large class of quantum master equations cannot be described by Lindblad dynamics. Here we assess the validity and the effectiveness of the quantum jump formalism when the dissipators in the quantum master equation describe a non-additive, open quantum system dynamics. We find that an additive unravelling of the non-additive quantum master equation does not generate a completely-positive dynamics in the scenario of perfect jump detection. Nevertheless, allowing for an imperfect jump detection scenario, we find that an additive unravelling is possible that reproduces the current and the fluctuations obtained via the Landauer-B\"uttiker formalism. 
\end{abstract}
                             
\maketitle

\section{Introduction}
\label{sec:introduction}
 
The study of quantum measurements plays a pivotal role in quantum information science, quantum many-body physics, and quantum thermodynamics.
Of particular importance are continuous measurements, which provide observable information in the form of an output current, represented by a classical, stochastic time series~\cite{Landi_2024}. The increasing focus on quantum science and technology has led to a burgeoning interest in the properties of these currents and their fluctuations in mesoscopic devices~\cite{Korotkov1999, *Korotkov2001, Goan2001, Murch2013, Koski2014, Haegele2018, Tilloy2018, Minev2019, Maillet2019, Barker2022, wadhia2025entropic, Fiusa2026}. On the one hand, current fluctuations fundamentally constrain measurement precision~\cite{Gammelmark2014, Cortez2017, Albarelli2018, Mihailescu2025, Khandelwal2025, Radaelli2026, Brattegard2026}; on the other, they dictate the stability of nanoscale machines. 
For example, quantum dots can function as autonomous thermoelectric engines~\cite{Benenti_2017} whose output is an electric current, fluctuating over time owing to the microscopic nature of the system. This and other examples highlight the importance of developing strategies to monitor these fluctuations, which enable the smooth operation of the device~\cite{Pietzonka_2018, Guarnieri_2019, Saryal_2021}.

In this regard, frameworks based on the Gorini-Kossakowski-Sudarshan-Lindblad (GKSL) quantum master equation (QME) have recently gained prominence. The formalism of quantum-jump trajectories~\cite{Ueda1990, Molmer1993, carmichael1993open} provides a useful instrument for modelling currents and fluctuations in open quantum systems~\cite{Landi_2024}, yielding a reliable experimental description of a host of atomic~\cite{Finn_1986,Reynaud_1988,Erber_1989,Carmichael_1989, Basche_1995, Peil_1999, Gleyzes_2007, Guerlin_2007} and solid-state systems~\cite{Jelezko_2002,Neumann_2010, Robledo_2011, Hatridge_2013, Sun_2014}. The precise description of quantum jumps is given by the concept of quantum operations and generalised measurements~\cite{nielsen2010quantum, benenti2019principles}, and is also referred to as an unravelling. The latter procedure is usually performed for QMEs in diagonal (Lindblad) form. This brings up the question whether we can use the formalism of quantum jump unravellings to evaluate current fluctuations when a GKSL master equation contains cross terms that are not in diagonal form? In
the context of GKSL QMEs, several works have employed quantum jump unravellings for evaluating current fluctuations in open quantum systems, establishing fluctuation relations~\cite{Manzano2018} and precision bounds~\cite{Vu_2022, Moreira_2025} and  multiple currents~\cite{Moreira_2025}, investigating the role of coherence~\cite{Miller2020,Prech_2023} or critical phenomena~\cite{Rota2018, Kewming2022, Matsumoto2025}, and studying regimes beyond weak system-environment coupling~\cite{Bettmann2024, Mahadeviya_2026}. However, the case of non-additive QMEs, involving cross terms due to the interference of multiple environments, 
has received much less attention.

Here we demonstrate the failure of quantum jump unravellings to characterise current fluctuations for an open quantum system in contact with two thermal reservoirs when the QME is non-additive. In particular, we consider two coupled single-level quantum dots~\cite{Mitchison_2018} each in contact with a thermal reservoir. This setup is an example of the class of resonant-level transport models~\cite{Caroli_1971, Meir_1992}, wherein a particle or heat current flows between the thermal baths under an applied temperature gradient or voltage bias and can, for example, represent a thermoelectric tunnel junction~\cite{Benenti_2017}, a thermal machine that generates entanglement~\cite{Brask_2015, Tavakoli2018heraldedgeneration} or a heat rectifier~\cite{Khomchenko_2022}. 

As was established in Ref.~\cite{Mitchison_2018}, the asymptotic dynamics of our system is governed by a time-local, non-additive master equation. Here, non-additivity means that there is no decomposition of the total dissipator into terms describing the action of each reservoir alone and independently. Such non-additive dynamics has been identified in various other open quantum systems under the influence of multiple environments~\cite{Kolodynski2018}, e.g., when phonon-induced dephasing competes with electronic transport~\cite{Giusteri2017, McConnell2019} or electromagnetic interactions~\cite{Maguire2019, Gribben2022}. Nevertheless, evaluating current fluctuations via the quantum-jump approach requires an \textit{additive} unravelling, where each jump is associated to a specific reservoir. 

Here we consider such additive unravellings, in both local and global bases, and investigate their validity in our non-additive dynamical setting. Specifically, we compare the predictions of these additive quantum-jump unravellings with the current statistics obtained via the Landauer-B\"uttiker (LB) approach. We find that, in general, the unravellings neither reproduce the LB results nor even generate a completely positive (CP) no-jump dynamics. The latter finding means that such a jump unravelling cannot be physically realised as a measurement process, even though it mathematically reproduces the correct density matrix on average. However, if one allows for \textit{imperfect detection}, there are specific parameter regimes where our additive quantum jump unravellings are both compatible with LB predictions and result in a CP no-click evolution. The interpretation is that non-additive dynamics is incompatible with strong continuous monitoring of each reservoir separately, as this destroys correlations between them. By contrast, only partially monitoring local exchanges of excitations can, in principle, allow non-additive interference between the two environments to persist.

This paragraph outlines the structure of our paper. Section~\ref{sec:model} reports the model of the system. Our methods for calculating steady-state observables, such as a particle current and its fluctuations, are given in Section~\ref{sec:approaches}. Analytical and numerical results demonstrating the break-down of the quantum jump approach for our setup are presented in Section~\ref{sec:unravelling}. In Section~\ref{sec:current_fluctuations}, we show the impact of the non-additive term in the QME on the current and the fluctuation comparing them with both the additive counterpart and the LB approach. Discussion and concluding remarks are drawn in Section~\ref{sec:discussion}. A list of appendices contains the technical details of our results.

\section{Model}
\label{sec:model}

\begin{figure}
    \centering
    \includegraphics[width=0.99\linewidth]{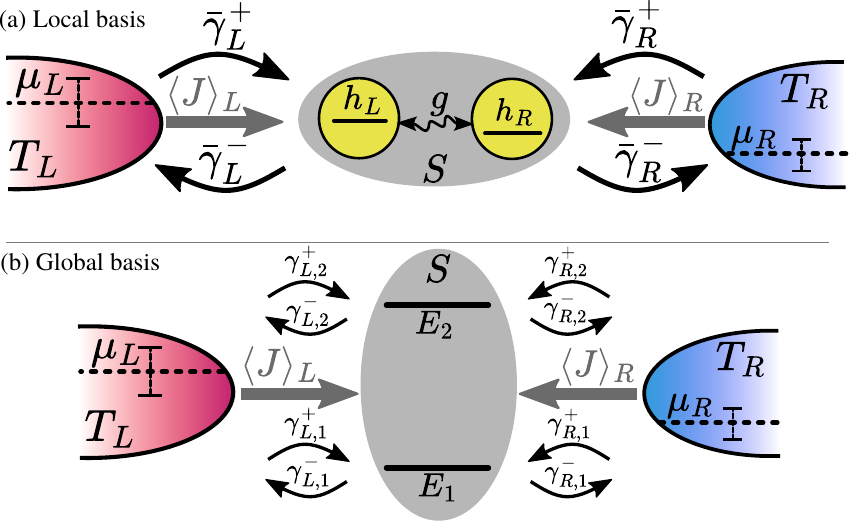}
    \caption{Double quantum dot connected to two fermionic reservoirs, denoted as a system $S$. Each dot contains only one energy level $h_{L,R}$ and the dots are coherently coupled with tunnel coupling $g$. Both dots can exchange particles independently with reservoirs ($L,R$) at temperatures $T_{L,R}$ and chemical potentials $\mu_{L,R}$, respectively. The double-dot quantum system is coupled to the two thermal baths, which ensures stationary-state  currents $\langle J \rangle_{L,R}$ flowing through the system. The upper panel sketches the setup in the local basis and the lower panel in the global basis. The rates $\bar{\gamma}^{\pm}_{\alpha}$, in the upper panel, characterise the jump processes between the reservoirs and the individual site $\alpha$; whereas, in the lower panel, the rates $\gamma^{\pm}_{\alpha, j}$ describe the jump processes between the reservoirs and the energy eigenstates of the Hamiltonian in Eq.~\eqref{eq:system_ham} of the open system.}
\label{fig:double_qauntum_dot_illustration}
\end{figure}

We consider a double quantum dot system $S$ in contact with the left (L) and right (R) reservoirs, as illustrated in the upper panel of Fig.~\ref{fig:double_qauntum_dot_illustration}. Each reservoir is characterised by a temperature $T_\alpha$ and a chemical potential $\mu_\alpha$, with $\alpha = L,R$. We set the Boltzmann's constant $k_B=1$ and the Planck's constant $\hbar =1$ throughout.

The Hamiltonian of the double quantum dot system $S$ is given by~\cite{Mitchison_2018}
\begin{equation}
    \hat{H}_S = \sum_{\alpha = L, R} h_\alpha \hat{n}_\alpha - \frac{g}{2} (\hat{c}_L^\dagger \hat{c}_R + \hat{c}_R^\dagger \hat{c}_L),
\label{eq:system_ham}
\end{equation}
where $ \hat{c}_\alpha$ is a fermionic annihilation operator acting on site $\alpha$ satisfying the anti-commutation relations $\{\hat{c}_\alpha, \hat{c}_{\alpha'}^\dagger\} = \delta_{\alpha \alpha'} $ and $ \{\hat{c}_\alpha, \hat{c}_{\alpha'}\} = 0 $. We denote the occupation of the quantum dot $\alpha$ by \( \hat{n}_\alpha = \hat{c}_\alpha^\dagger \hat{c}_\alpha \). The local energies $h_\alpha$ are parametrised by their mean $h = \frac{1}{2}(h_L+h_R)$ and the detuning $\delta= h_L-h_R$, with $g$ being the coherent coupling between the dots.

The total Hamiltonian can be written as, 
\begin{equation}
\hat{H}=\hat{H}_S+\hat{H}_B+\hat{H}_{SB},
\label{eq:total_ham}
\end{equation}
where $\hat{H}_B$ represents the environment Hamiltonian (left and right reservoirs) and $\hat{H}_{SB}$ accounts for the interaction between system and environment (see Appendix ~\ref{appsec:Reservoirs} for details)~\cite{Mitchison_2018}. 

An exact time-local master equation describing the dynamics of $S$ has been derived for arbitrary spectral densities~\cite{Mitchison_2018, Jin_2010, Yang_2013}. We follow Ref.~\cite{Mitchison_2018} and, for the sake of completeness, we summarise in this Section and in Appendix~\ref{appsec:derivation}, the main points of the derivation.  In our work, we are interested in the asymptotic dynamics ($t \rightarrow \infty$) of the system, we specialise to the Newns spectral density~\cite{Newns_1969}, and consider the system-reservoir tunnelling energies $\sqrt{\Gamma \Omega}$ to be much smaller than the intra-reservoir tunnelling energies $\Omega$ (see Appendix~\ref{appsec:Reservoirs} and \ref{appsec:dissipator} for details).
The asymptotic dynamics can then be described by the following master equation~\cite{Mitchison_2018},
\begin{equation}
\partial_t \hat{\rho}_S(t) = -\text{i} [\hat{H}_S, \hat{\rho}_S(t)] + \mathcal{L}\hat{\rho}_S(t) \equiv \mathfrak{L}\hat{\rho}_S(t),
\label{eq:exact_master_equation}
\end{equation} 
where $\mathfrak{L}$ denotes the  Liouvillian, and the total dissipator $\mathcal{L}$ can be decomposed as follows
\begin{eqnarray}
 \mathcal{L} = \mathcal{L}_L^{\rm c}+ \mathcal{L}_R^{\rm c}+\mathcal{L}_{LR}^{\rm c}.
\label{eq:decompositions}     
\end{eqnarray}
Here, the local dissipators are represented by Lindblad generators in the site basis acting on a single site $\alpha$, 
\begin{equation}
\mathcal{L}_\alpha^{\rm c}  = \bar{\gamma}_\alpha^- \mathcal{D}[\hat{c}_\alpha] + \bar{\gamma}_\alpha^+ \mathcal{D}[\hat{c}_\alpha^\dagger],
\label{eq:generatorlocal_s}  
\end{equation}
with decay and gain rates $\bar{\gamma}^{\pm}_{\alpha}$ derived in Appendix~\ref{appsec:dissipator} and Lindblad dissipators given by $\mathcal{D}[\hat{c}]~\bullet = \hat{c}\bullet\hat{c}^{\dagger} - \frac{1}{2}\{ \hat{c}^{\dagger}\hat{c},\bullet\}$. 
The contribution $\mathcal{L}_{LR}^{\rm c} $ corresponds to delocalised processes involving both sites together,
\begin{widetext}
\begin{equation}
\mathcal{L}_{LR}^{\rm c}\hat{\rho}_S = \sum_{\alpha \neq \alpha'} \left[ \bar{\Lambda}_{\alpha \alpha'}^+ \left( \hat{c}_{\alpha'}^\dagger  \hat{\rho}_S \hat{c}_\alpha - \frac{1}{2} \{ \hat{c}_\alpha \hat{c}_{\alpha'}^\dagger, \hat{\rho }_S \} \right) + \bar{\Lambda}_{\alpha \alpha'}^- \left( \hat{c}_{\alpha} \hat{\rho }_S \hat{c}_{\alpha'}^\dagger - \frac{1}{2} \{ \hat{c}_{\alpha'}^\dagger \hat{c}_\alpha, \hat{\rho }_S \} \right) \right],
\label{eq:interference_local}  
\end{equation}    
\end{widetext}
with the rates ${\bar{\Lambda}}_{LR}^{\pm}=({\bar{\Lambda}}_{RL}^{\pm})^*$ given in Appendix~\ref{appsec:dissipator}.  

It may be convenient to analyse the problem in a basis that diagonalises $\hat{H}_S$, which is related to the site basis by the orthogonal rotation $R$, such that 
$\hat{\mathbf{d}}= (\hat{d}_1, \hat{d}_2)^{\text{T}} = \text{R} \hat{\mathbf{c}} = \text{R} (\hat{c}_1, \hat{c}_2)^{\text{T}}$ (see Appendix~\ref{appsec:Reservoirs} for details). In this case, the total dissipator reads
\begin{eqnarray}
 \mathcal{L} = \mathcal{L}_L^{\rm d}+ \mathcal{L}_R^{\rm d}+\mathcal{L}_{LR}^{\rm d},
\label{eq:decompositions_d}     
\end{eqnarray}
where the Lindblad generators $\mathcal{L}_\alpha^{\rm d}$ have the form
\begin{equation}
\mathcal{L}_{\alpha}^{\rm d} = \sum_{j=1}^{2} \left( \gamma_{\alpha,j}^{-} \mathcal{D}[\hat{d}_j] + \gamma_{\alpha,j}^{+} \mathcal{D}[\hat{d}_j^{\dagger}] \right),
\label{eq:generators_global}
\end{equation}
with $\gamma_{\alpha,j}^{-}$, $\gamma_{\alpha,j}^{+}$ given in Appendix~\ref{appsec:dissipator}.
The interference term $\mathcal{L}_{LR}^{\rm d}$, in turn, can be written as
\begin{widetext}
\begin{equation}
\mathcal{L}_{LR}^{\rm d}\hat{\rho}_S =\sum_{j \neq k} \Lambda^+_{jk} \left[ \left(\hat{d}_k^\dagger \hat{\rho}_S \hat{d}_j - \frac{1}{2} \{\hat{d}_j \hat{d}_k^\dagger, \hat{\rho}_S\} \right) - \left(\hat{d}_j \hat{\rho}_S \hat{d}_k^\dagger - \frac{1}{2} \{\hat{d}_k^\dagger \hat{d}_j, \hat{\rho}_S\}\right) \right],
\label{eq:interference_global}
\end{equation}
\end{widetext}
where the rates satisfy ${\Lambda}_{12}^{+}=({\Lambda}_{21}^{+})^*$ (see Appendix~\ref{appsec:dissipator} for details).
In this case, the dissipator $\mathcal{L}_\alpha^{\rm d}$ is a Lindblad generator describing the thermalising effect of the individual bath $\alpha$ with system $S$, while the interference term, $\mathcal{L}_{LR}^{\rm d}$ appears whenever the left and right baths are not in equilibrium with each other, as can be seen from Eq.~\ref{eq:interference_re_im_parts} in Appendix~\ref{Appendix_B1} and generates coherence in the basis of $\hat{H}_S$.

 The expressions in Eqs.~\eqref{eq:interference_local} and \eqref{eq:interference_global} constitute the cross terms that yield a non-additive dynamics. There exist low-temperature parameter regimes where these cross terms even result in asymptotically non-Markovian dynamics~\cite{Hall_2014}, meaning that the evolution is not positive-divisible~\cite{Breuer2016} and therefore cannot be written in Lindblad form~\cite{Mitchison_2018}. However, there exists a significant parameter regime in which the dynamics generated by Eq.~\eqref{eq:exact_master_equation} is CP-divisible and can be written in canonical Lindblad form
\begin{equation}
\partial_t \hat{\rho}_S(t) = -\text{i} [\hat{H}_S, \hat{\rho}_S(t)] + \sum_k \mathcal{D}[\hat{L}_k]\hat{\rho}_S(t),
\label{eq:diaganal_form}
\end{equation}
where $\hat{L}_k$ are jump operators obtained by diagonalising the dissipator  $\mathcal{L}$~\cite{Breuer_2007}. In this form, the jump operators $\hat{L}_k$ are explicitly non-additive, i.e. they describe a superposition of processes mediated by the left and right bath, in both the site (c) or eigenenergy (d) representation. 

\section{Current fluctuations: Landauer-B\"uttiker and quantum jump approaches}
\label{sec:approaches}

The  nonequilibrium set-up described in the previous Section supports a particle current due to temperature and chemical potential imbalance between the reservoirs. The average current and fluctuations can be calculated by employing the Landauer-B\"uttiker (LB)~\cite{Buttiker_1986} formalism, which provides  expressions for these quantities for non-interacting fermions in the asymptotic limit~\cite{Prech_2023}.
In this way, the net average particle current exchanged with reservoir $\alpha$ can be written as~\cite{datta2005quantum, di2008electrical} 
\begin{eqnarray}
  \langle J \rangle_{L/R} = \int_{-\infty}^{\infty}  \mathcal{T}(\omega)[f_{L/R}(\omega)-f_{R/L}(\omega)]~\mathrm{d}\omega,  
  \label{eq:particle_current}
\end{eqnarray} 
where $f_\alpha(\omega) = [e^{(\omega - \mu_\alpha)/T_\alpha}+1]^{-1}$ is the Fermi–Dirac distribution in reservoir $\alpha$ and $\mathcal{T}(\omega)$ is the transmission function of the form 
\begin{equation}
\mathcal{T}(\omega)=\frac{\pi g^2}{2} \varphi_1(\omega) \varphi_2(\omega),
\label{eq:trasmission_function}
\end{equation}
where $\varphi_j(\omega)$ characterises the distribution of energies in the bath (further details given in Appendix~\ref{appsec:dissipator})
and $ 0 \leq \mathcal{T}(0) \leq 1$~\cite{Agarwalla_2018}.
Additionally, the fluctuations of the same net current can be expressed as~\cite{Blanter_2000,Prech_2023} 
\begin{widetext}
\begin{equation}
    \langle \langle J^{2} \rangle \rangle_{L/R}= \int_{-\infty}^{\infty} \mathcal{T}(\omega) \{ [f_{L/R}(\omega)+f_{R/L}(\omega)-2f_{L/R}(\omega)f_{R/L}(\omega)]-\mathcal{T}(\omega)[f_{L/R}(\omega)-f_{R/L}(\omega)]^2\}
    ~\mathrm{d}\omega. 
\label{eq:current_fluctuations}
\end{equation} 
\end{widetext} 
This expression accounts for the equilibrium noise and for the non-equilibrium (or shot) noise. 
In the LB approach, the net charge transferred from the system into the left or right reservoir is obtained by integrating the DC current measured by an
ammeter over a certain time interval. Theoretically, this can be modelled as a two-point measurement of the number of
electrons in a given reservoir~\cite{Esposito_2009}.
 
We note that whenever the system evolution is described by a GKSL master equation in Lindblad form, current fluctuations can be assessed by employing the \emph{quantum-jump approach}~\cite{Landi_2024}. This is typically done by writing down the diagonal master equation~\eqref{eq:diaganal_form} and assigning weights $\nu_k$ to each of the jump operators $\hat{L}_k$. However, in the case of non-additive dynamics, governed by Eq.~\eqref{eq:exact_master_equation} this approach is not applicable since $\hat{L}_k$ cannot be associated to one bath or another, while our goal is to calculate the current flowing into a single bath. As a result, we cannot choose weights $\nu_k$ such that the current and its fluctuations can be calculated. 

Therefore, let us instead consider \emph{additive} jump operators associated with the exchange of particles with reservoir $\alpha$, where now each jump operator $\hat{L}_k$ (to be defined explicitly in the next section) will represent either the creation or annihilation of a fermion in the system mediated by reservoir $\alpha$. To construct the net current flowing into the system from reservoir $\alpha$ into the system, we associate the weight $\nu^\alpha_k = +1$ with creation operators associated with reservoir $\alpha$, $\nu^\alpha_{k} = -1$ with annihilation operators associated with reservoir $\alpha$, while we set $\nu^\alpha_k=0$ for all other jump operators. In the steady state $\hat{\rho}_{\mathrm{ss}} = \lim_{t \rightarrow \infty} \hat{\rho}_S(t)$, the average current $\langle J \rangle_\alpha$ and current fluctuations $\langle \langle J^2 \rangle \rangle_{\alpha}$ obtained by using the quantum jump approach in vectorised notation (see Appendix~\ref{appsec:Vectorisation} for more details), are then given by~\cite{Landi_2024}
\begin{eqnarray}
&\langle J \rangle_{\alpha}& = \langle \langle \mathbf{1} | \mathcal{J}_\alpha | \hat{\rho}_{\mathrm{ss}} \rangle \rangle
=\sum_{k} \nu^\alpha_k \mathrm{Tr}\{\hat{L}_k^{\dagger} \hat{L}_k \hat{\rho}_{\mathrm{ss}}\}, \label{average_jump_current}\\
&\langle \langle J^2 \rangle \rangle_{\alpha}& = K_{\alpha}
- 2 \langle \langle \mathbf{1} | \mathcal{J}_\alpha\mathcal{L}^{+}\mathcal{J}_\alpha | \hat{\rho}_{\mathrm{ss}} \rangle \rangle. \label{current_fluctuations_eq_tot}
\end{eqnarray}
These definitions are written in terms of the current superoperator $\mathcal{J}_\alpha$, whose action on a density matrix $\rho$ yields $\mathcal{J} (\hat{\rho}) = \sum_{k} \nu^\alpha_k \hat{L}_k \hat{\rho}\hat{L}_k^{\dagger}$. The quantity $K_\alpha = \sum_{k} (\nu^{\alpha}_k)^2 \mathrm{Tr} \{ \hat{L}_k^{\dagger} \hat{L}_k \hat{\rho}_{\mathrm{ss}} \}$ is the dynamical activity associated with reservoir $\alpha$~\cite{DiTerlizzi_2019}, the symbol $\mathcal{L}^+$ denotes the Drazin pseudoinverse operator, whose properties and definition are discussed in Ref.~\cite{Landi_2024}, and $\mathbf{1}$ is the identity operator.

In the next Section, we investigate whether the formalism of Eqs.~\eqref{average_jump_current} and \eqref{current_fluctuations_eq_tot} can be used to assess current fluctuations in the non-additive QME dynamics by considering both perfect and imperfect detection schemes.

\section{Investigating the quantum jump approach in the non-additive dynamics}
\label{sec:unravelling}

The non-additive master equation in Eq.~\eqref{eq:exact_master_equation} provides a completely positive and trace preserving (CPTP) description of the system dynamics at the ensemble level. By  unravelling the QME for evaluating current fluctuations one implicitly assumes that the quanta exchanged, for example with heat baths or with specific sites, can be monitored.

Most generally, detection processes are theoretically described by
CPTP quantum instruments acting on the system state $\hat{\rho}_S$ as~\cite{wiseman2009quantum, Landi_2024}
\begin{equation}
    \mathcal{E}(\hat{\rho}_S)= \sum_k  \mathcal{E}_k (\hat{\rho}_S) =  \sum_k \hat{M}_k \hat{\rho}_S \hat{M}_k^\dagger,
\label{eq:kraus_decomposition}
\end{equation}
where $\mathcal{E}_k$ are superoperators, and $\hat{M}_k$ are Kraus operators satisfying $\sum_k \hat{M}_k^\dagger \hat{M}_k = 1$.
 
For the non-additive Liouvillian $\mathfrak{L}$ in Eq.~\eqref{eq:exact_master_equation}, the system's evolution can be written as $\hat{\rho}_S(t) = e^{\mathfrak{L}t}\hat{\rho}_S(0)$. Given an infinitesimal time interval $dt$, we can expand the system's evolution as
\begin{eqnarray}
\hat{\rho}_S(t+ dt) &= e^{\mathfrak{L} dt}\hat{\rho}_S(t) 
 = \hat{\rho}_S(t) +  dt \mathfrak{L} ( \hat{\rho}_S(t) ) + \mathcal{O}(dt^2) \nonumber \\  
&=\mathcal{E}(\hat{\rho}_S(t)) +\mathcal{O}(dt^2).
\label{eq:cptp_decomposition}
\end{eqnarray}
We note that such unravellings of the dynamics are not unique, i.e., there are infinitely many possible decompositions of the system dynamics in terms of Kraus operators.

In the following, we consider specific \emph{additive} unravellings associated with monitoring jumps in and out single sites (i.e. in the local basis) or in and out the heat baths (i.e. in the energy or global basis), and investigate whether they are compatible with the monitoring or detection of jumps in the non-additive dynamics. Measuring in two distinguished bases corresponds to different methods of counting fermions: Detecting fermions in the local basis enables a detector to know from which site the fermion originates; on the other hand, in the global basis, a detector cannot distinguish fermions originating from the left site from those coming from the right site.

\subsection{Perfect Detection}
\label{perfect_detection}
Suppose we can perfectly detect all jumps between the system and the baths when monitoring quanta in and out of the left ($L$) or right ($R$) sites, as illustrated in Fig~\ref{fig:double_qauntum_dot_illustration}~(a). 
In this case, the jumps are represented by the jump operators $\hat{L}_1 = \sqrt{\bar{\gamma}^{-}_{L}} \hat{c}_L$, $\hat{L}_2 = \sqrt{\bar{\gamma}^{+}_{L}} \hat{c}^\dagger_L$, $\hat{L}_3 = \sqrt{\bar{\gamma}^{-}_{R}} \hat{c}_R$, and $\hat{L}_4 = \sqrt{\bar{\gamma}^{+}_{R}} \hat{c}^\dagger_R$ within the dissipators $\mathcal{L}_\alpha^{\rm c}$.
We can therefore identify the maps $\mathcal{E}_k^{\rm c}$, for $k = 1, \ 2, \ 3, \ 4$, as jump superoperators, which can be written as
\begin{eqnarray}
\label{eq:cptp_maps}
\mathcal{E}_k^{\rm c}(\hat{\rho}_S) = dt \ \hat{L}_k \hat{\rho}_S \hat{L}_k^\dagger.  
\end{eqnarray}
We note that all expressions in Eq.~\eqref{eq:cptp_maps} are completely positive (CP) maps.
By replacing the maps $\mathcal{E}_k^{\rm c} $ in Eq.~\eqref{eq:cptp_decomposition}, we find that
\begin{equation}
\mathcal{E}_0^{\rm c} = \mathcal{I} +   \mathfrak{L}dt  - \sum_{k=1}^4 \mathcal{E}_k^{\rm c}  + \mathcal{O} (dt^2).
\label{eq:final_form}
\end{equation}
Thus, assuming we can monitor the local jumps, at each interval of time $dt$, the state of the system is updated as
\begin{equation}
    \hat{\rho}_S \mapsto \mathcal{E}_k^{\rm c}(\hat{\rho}_S),
\end{equation}
when a jump in the dissipation channel $k$ is observed. If no jump is recorded, in turn,
\begin{equation}
    \hat{\rho}_S \mapsto \mathcal{E}_0^{\rm c}(\hat{\rho}_S).
\end{equation}
In other words, the no-jump evolution after each infinitesimal time interval $dt$ is governed by the map $\mathcal{E}_0^{\rm c}$. 
At the level of a single trajectory, the no-jump evolution between the observation of jumps is obtained through the repeated application of $\mathcal{E}_0^{\rm c}$. In this way, we can show that the map which will govern the evolution between jumps for arbitrary, stochastic time intervals $\Delta t$ between two consecutive jumps, is given by
\begin{equation}\label{Mc}
    \mathcal{M}_0^{\rm c} = e^{(\mathcal{L}_{LR}^{\rm c}-i \mathcal{H}_{\text {eff}}^{\rm c})\Delta t},
\end{equation}
where $\Delta t$ is an arbitrary time interval between consecutive jumps and $\mathcal{H}_{\text{eff}}^{\rm c}~\bullet \equiv  \hat{H}_{\text {eff}}^{\rm c}~\bullet -  \bullet~\hat{H}_{\text {eff}}^{\rm c \dagger}$ is a superoperator, with $\hat{H}_{\text {eff}}^{\rm c} = \hat{H}_S - \frac{i}{2}\sum_{k=1}^4 \hat{L}_k^\dagger \hat{L}_k$.

Next, we assume that we can monitor the exchange of quanta between the double-dot system and the heat baths, as depicted in Fig~\ref{fig:double_qauntum_dot_illustration} (b).
In this case, we have the following jump operators in the global basis, $\hat{L}_1 = \sqrt{\gamma^{-}_{L,1}} \hat{d}_1$, $\hat{L}_2 = \sqrt{\gamma^{+}_{L,1}} \hat{d}^\dagger_1$, $\hat{L}_3 = \sqrt{\gamma^{-}_{L,2}} \hat{d}_2$, $\hat{L}_4 = \sqrt{\gamma^{+}_{L,2}} \hat{d}^\dagger_2$, $\hat{L}_5 = \sqrt{\gamma^{-}_{R,1}} \hat{d}_1$, $\hat{L}_6 = \sqrt{\gamma^{+}_{R,1}} \hat{d}^\dagger_1$, $\hat{L}_7 = \sqrt{\gamma^{-}_{R,1}} \hat{d}_2$, and $\hat{L}_8 = \sqrt{\gamma^{+}_{R,2}} \hat{d}^\dagger_2$ within the dissipators $\mathcal{L}^{\rm d}_\alpha$. The corresponding jump superoperators $\mathcal{E}_p^{\rm d}$, for $p = 1, \ldots, 8$, are therefore given by
\begin{eqnarray}
\label{eq:cptp_maps_global}
\mathcal{E}_p^{\rm d}(\hat{\rho}_S) = dt \ \hat{L}_p \hat{\rho}_S \hat{L}_p^\dagger.
\end{eqnarray}
We proceed in the same way as in the local basis to determine the no-jump evolution $\mathcal{E}_0^{\rm d}$, 
\begin{eqnarray}
    \mathcal{E}_0^{\rm d}  
= \mathcal{I} +   \mathfrak{L}dt  - \sum_{p=1}^8 \mathcal{E}_p^{\rm d}  + \mathcal{O} (dt^2).
\label{eq:map_extended_form_gloabl}
\end{eqnarray}
Then, we get that the evolution between jumps in the energy eigenbasis is given by the following map \begin{equation}\label{Md}
    \mathcal{M}_0^{\rm d} = e^{(\mathcal{L}_{LR}^{\rm d}-i \mathcal{H}_{\text {eff}}^{\rm d})\Delta t},
\end{equation}
with $\mathcal{H}_{\text{eff}}^{\rm d}~\bullet \equiv  \hat{H}_{\text {eff}}^{\rm d}~\bullet  -  \bullet~\hat{H}_{\text {eff}}^{\rm d \dagger}$ and $\hat{H}_{\text {eff}}^{\rm d} = \hat{H}_S - \frac{i}{2}\sum_{k=1}^8 \hat{L}_k^\dagger \hat{L}_k$. 
To verify if these unravellings in the local or global bases are associated with physical measurement descriptions, we need to verify if the maps in Eqs.~\eqref{Mc} and~\eqref{Md} are CP, respectively~\cite{bengtsson2017geometrys}. With this aim, we employ the Choi's criterion, which states that the Choi matrix of CP quantum map must be positive~\cite{bengtsson2017geometrys}. Owing to a large size of the Choi matrix for our system ($16 \times 16$), we perform numerical tests to ascertain whether the maps are CP. 
The results of these tests, which we detail in Appendix~\ref{appsec:no_jump},  show that neither $\mathcal{M}_0^{\rm c}$ nor $\mathcal{M}_0^{\rm d}$ correspond to a CP evolution between jumps. This implies that perfectly detecting jumps as described by the additive unravellings considered here, while preserving the full non-additive dynamics generated by the dissipators in Eqs.~\eqref{eq:interference_local} and~\eqref{eq:interference_global}, is impossible. To summarise this Section's result, the cross terms $\mathcal{L}^{\rm m}_{LR}$ lead to the breakdown of the quantum-jump approach: These terms do not generate a positive evolution by themselves~\cite{Mitchison_2018}. Furthermore, although the terms, such as $\bar{\Lambda}_{\alpha \alpha'}^+ \hat{c}_{\alpha'}^\dagger \hat{\rho}_S \hat{c}_\alpha$, resemble local jump terms, for example, $\bar{\gamma}_{\alpha}^+ \hat{c}_{\alpha} \hat{\rho}_S \hat{c}_\alpha^\dagger$ ($\bar{\gamma}_{\alpha}^- \hat{c}_{\alpha}^\dagger \hat{\rho}_S \hat{c}_\alpha$), they are non-local terms and do not represent quantum jumps because they map a density matrix to a non-physical state, and thus cannot be considered as physical processes.

In this way, the immediate practical implication of our results is that in the case of perfect detection, \emph{current fluctuations in the non-additive dynamics cannot be computed by assuming additive quantum jump unravellings in the local or global basis}.

\subsection{Imperfect Detection}

Here, we consider the case of imperfect detection, meaning that some jumps go undetected. Our goal is to investigate whether additive unravellings of the non-additive evolution in imperfect detection situations
can be described by CPTP quantum instruments, thereby allowing us to obtain the average current and fluctuations given by the LB formalism in Eqs.~\eqref{eq:particle_current} and~\eqref{eq:current_fluctuations}.

Let us assume that each jump in an unravelling is measured with an efficiency $\eta_k^{\rm c} \in [0,1]$, such that $\eta_k^{\rm c} = 1$ correspond to the perfect efficiency~\cite{Landi_2024}.
We first consider unravellings in the site basis, as defined by the following jump operators 
$\hat{L}_1^1 = \sqrt{\bar{\gamma}^{-}_L\eta_1^{\rm c}} \hat{c}_L$, $\hat{L}_2^1 = \sqrt{\bar{\gamma}^{+}_L\eta_2^{\rm c}} \hat{c}_L^\dagger$, $\hat{L}_3^1 = \sqrt{\bar{\gamma}^{-}_R\eta_3^{\rm c}} \hat{c}_R$, and $\hat{L}_4^1 = \sqrt{\bar{\gamma}^{+}_R\eta_4^{\rm c}} \hat{c}_R^\dagger$, with $\eta_k^{\rm c} = \tilde{\gamma}^{\pm}_\alpha/\bar{\gamma}^{\pm}_\alpha$. The new rates $\tilde{\gamma}_\alpha^{\pm}$ are determined such that the LB average current and fluctuations from Eqs.~\eqref{eq:particle_current} and~\eqref{eq:current_fluctuations} are matched.
Specifically, the rates are derived from a system of equations given by Eqs.~\eqref{average_jump_current} and~\eqref{current_fluctuations_eq_tot}, the detailed derivation of which is discussed in Appendix~\ref{appsec:positivity_complete_positivity}. 
Using this, we can follow the same procedure as in Subsection \ref{perfect_detection} to construct a new map $\mathcal{M}_0^{\rm c, j}$ of the form
\begin{equation}
\label{Mc_Jimp}
    \mathcal{M}_0^{\rm c, j} = e^{ ( \mathcal{L}_{\rm j}^{\rm c}+\mathcal{L}_{LR}^{\rm c}-i \mathcal{H}_{\text{eff}}^{\rm c})\Delta t},
\end{equation} 
where
\begin{widetext}
\begin{align}
    \mathcal{L}_{\rm j}^{\rm c} \hat{\rho}_S = \bar{\gamma}^{-}_{L}(1 - \eta_1^{\rm c}) \hat{c}_L \hat{\rho}_S \hat{c}_L^\dagger + \bar{\gamma}^{+}_{L}(1 -\eta_2^{\rm c}) \hat{c}_L^\dagger \hat{\rho}_S \hat{c}_L + \bar{\gamma}^{-}_{R}(1  - \eta_3^{\rm c}) \hat{c}_R \hat{\rho}_S \hat{c}_R^\dagger +\bar{\gamma}^{+}_{R}(1  - \eta_4^{\rm c}) \hat{c}_R^\dagger \hat{\rho}_S  \hat{c}_R ,
\end{align}
\end{widetext}
with rates $\bar{\gamma}^{\pm}_{\alpha} - \tilde{\gamma}^{\pm}_{\alpha}>0$. Note that Eq.~\eqref{Mc_Jimp} has an extra term in the argument of the exponential, $\mathcal{L}_{\rm j}^{\rm c} \Delta t$, when compared to Eq.~\eqref{Mc}, which arises
from the undetected local jumps, such as, $\hat{L}_1^0 = \sqrt{\bar{\gamma}^{-}_{L}(1 -\eta_1^{\rm c}) } \hat{c}_L$, $\hat{L}_2^0 = \sqrt{\bar{\gamma}^{+}_{L}(1 -\eta_2^{\rm c})} \hat{c}^\dagger_L$,  $\hat{L}_3^0 = \sqrt{\bar{\gamma}^{-}_{R}(1 -\eta_3^{\rm c}) } \hat{c}_R$, and $\hat{L}_4^0 = \sqrt{\bar{\gamma}^{+}_{R}(1 -\eta_4^{\rm c}) } \hat{c}^\dagger_R$,  with positive rates $\bar{\gamma}^{\pm}_{\alpha} - \tilde{\gamma}^{\pm}_{\alpha} >0$. The new dissipator $\mathcal{L}_{\rm j}^{\rm c}$ accounts for the partial detection of local jumps. In this way, the superoperator $\mathcal{L}_{\rm j}^{\rm c}+\mathcal{L}_{LR}^{\rm c}-i \mathcal{H}_{\text{eff}}^{\rm c}$ describes the \emph{no-click} evolution. The no-click evolution includes the no-jump evolution, as well as the undetected jumps, and it gives the state update rule when fermions are not registered on a measuring device. 

Note that attempting to unravel the non-additive master equation~\eqref{eq:exact_master_equation} in the global basis using eight \emph{additive} jump operators, $\hat{L}_1^1 = \sqrt{\gamma^{-}_{L,1} \eta^d_1} \hat{d}_1$, $\hat{L}_2^1 = \sqrt{\gamma^{+}_{L,1}\eta^d_2} \hat{d}^\dagger_1$, $\hat{L}_3^1 = \sqrt{\gamma^{-}_{L,2}\eta^d_3} \hat{d}_2$, $\hat{L}_4^1 = \sqrt{\gamma^{+}_{L,2}\eta^d_4} \hat{d}^\dagger_2$, $\hat{L}_5^1 = \sqrt{\gamma^{-}_{R,1}\eta^d_5} \hat{d}_1$, $\hat{L}_6^1 = \sqrt{\gamma^{+}_{R,1}\eta^d_6} \hat{d}^\dagger_1$, $\hat{L}_7^1 = \sqrt{\gamma^{-}_{R,1}\eta^d_7} \hat{d}_2$, and $\hat{L}_8^1 = \sqrt{\gamma^{+}_{R,2}\eta^d_8} \hat{d}^\dagger_2$, with $\eta^d_p = \tilde{\Gamma}^\pm_{\alpha, j}/\gamma^{\pm}_{\alpha, j}$ yields an underdetermined system, because the eight rates $\tilde{\Gamma}^\pm_{\alpha, j}$ cannot be \emph{uniquely determined} from only four equations -- two systems of  Eqs.~\eqref{average_jump_current} and~\eqref{current_fluctuations_eq_tot} for left and right quantities -- as detailed in Appendix~\ref{appsec:positivity_complete_positivity}.

\begin{figure}[!t]
     \centering
     \includegraphics[width=1.\linewidth]{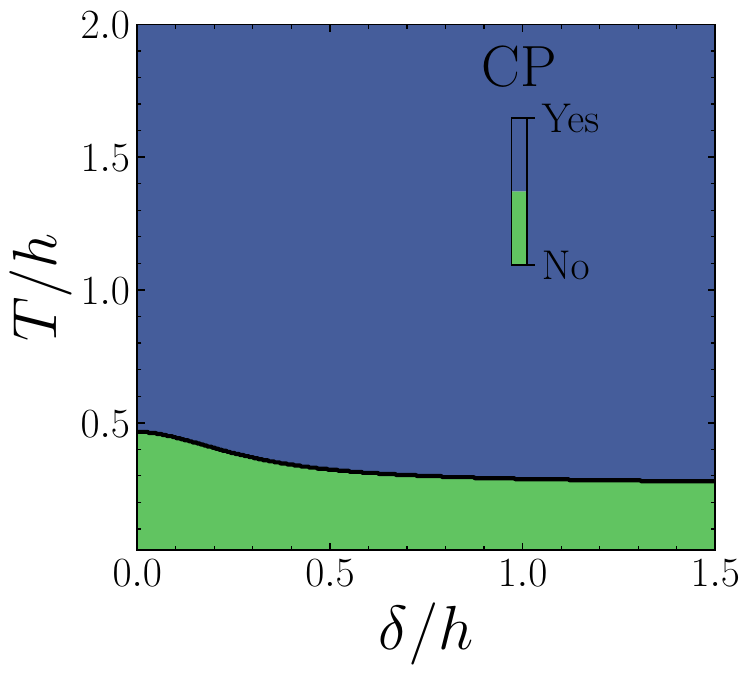}
     \caption{Complete positivity (CP) of the quantum map $\mathcal{M}_0^{\rm c,j}$ in the site basis as a function of the local energy levels detuning $\delta/h$ and temperature $T = T_L=T_R$. Parameters: $h=1$, $g=0.5$, $\Gamma = 0.02 $, $\Omega = 100$, $\mu_L/h = 0.7$, $\mu_R/h=0.0$. The blue colour denotes that the quantum map is CP and the obtained rates $\tilde{\gamma}^{\pm}_{\alpha}$ are smaller than the actual rates $\bar{\gamma}^{\pm}_{\alpha}$, i.e. $\bar{\gamma}^{\pm}_{\alpha} > \tilde{\gamma}^{\pm}_{\alpha}$ (``Yes"), while the green one corresponds to a non-CP quantum map (``No") with positive or negative obtained rates. This plot is invariant with respect to time between jumps $\Delta t$, which we verified for $1001$ values of $\Delta t \in [10^{-3},10^3]$.} 
     \label{fig:cpness_of_imd}
\end{figure}

We are now in a position to verify if the map in Eq.~\eqref{Mc_Jimp} is CP. Once again, we employ the Choi criterion. The results are shown in a contour plot in Fig.~\ref{fig:cpness_of_imd}. We see that $\mathcal{M}_0^{\rm c, j}$ is CP for sufficiently high temperatures. This implies that, in these parameter regimes, the additive unravelling of the non-additive master equation~\eqref{eq:exact_master_equation} can be used to assess current fluctuations. Conversely, at low temperatures, $T \lesssim h$, the $\mathcal{M}_0^{\rm c, j}$ is generally not CP. Indeed, in this parameter regime, the cross term $\mathcal{L}^{\rm c}_{LR}$ is non-negligible, as the rates  $\bar{\Lambda}_{LR}^\pm$ ($\bar{\Lambda}_{RL}^\pm$) in Eq.~\eqref{eq:interference_local} are comparable with respect to the single-site rates $\bar{\gamma}_{L,R}^\pm$ (see Appendix~\ref{appsec:positivity_complete_positivity} for details).

In conclusion, although imperfect detection schemes do not generally lead to CP no-click maps, there is a large parameter regime where the no-click maps are CP and, hence, an additive unravelling of the non-additive QME matching the current, and its fluctuation, predicted by the LB formalism is possible.  We also note that the parameter regimes where the no-click maps are CP can overlap with regions where the non-additive evolution is non-Markovian, as illustrated in Appendix~\ref{appsec:non_markovianity}, thus allowing for additive unravelling of non-Markovian QMEs. 

\section{Current fluctuations in the non-additive dynamics} 
\label{sec:current_fluctuations} 

 \begin{figure}[!t]
    \centering
    \includegraphics[width=1.0\linewidth]{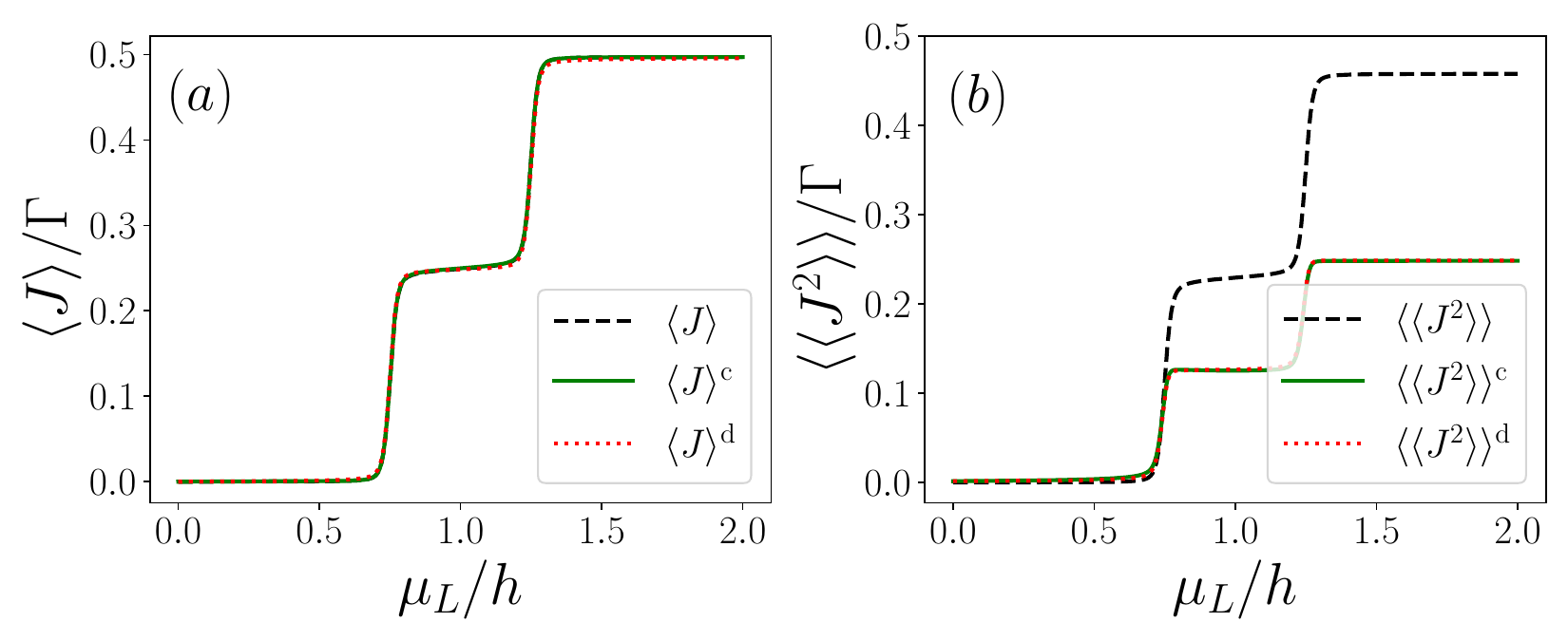}
    \caption{Average steady-state current $\langle J \rangle/\Gamma$ (a) and its fluctuations $\langle \langle J^2 \rangle \rangle/\Gamma$ (b) through the double quantum dot with the variable chemical potential of the left reservoir $\mu_L/h$. The the subscript $\text{c}$ and $\text{d}$ refers to the site and energy eigenbasis, respectively. Parameters: $h=1$, $\delta =0.0$, $g =0.5$, $\Delta = \sqrt{\delta^2+g^2}$, $\Gamma = 0.02 \Delta$, $\Omega = 10$, $T_L=T_R =0.02 \Delta$, $\mu_R/h=0$.}
    \label{fig:current_fluctuations_vs_mu} 
\end{figure}

In order to illustrate our result that current fluctuations cannot be assessed by additive jump unravellings in the local or global basis in the case of perfect detection, we compare the average left current and its fluctuations calculated using  Eqs.~\eqref{average_jump_current} and~\eqref{current_fluctuations_eq_tot}, to their values computed using the LB approach.
As the non-additivity of the asymptotic dynamics can be controlled by tuning the chemical potentials of the reservoirs~\cite{Mitchison_2018},
we compare these quantities in Fig.~\ref{fig:current_fluctuations_vs_mu}  as a function of the chemical potential of the left reservoir $\mu_L/h$, while the chemical potential of the right reservoir is set to 0.

In Fig.~\ref{fig:current_fluctuations_vs_mu}(a), we plot the LB average current, $\langle J\rangle$, as well as $\langle J\rangle^{\text{c}}$ and $\langle J\rangle^{\text{d}}$. We clearly see that $\langle J\rangle^{c}$ and $\langle J\rangle^{\text{d}}$ almost coincide with the LB average current $\langle J\rangle$. However, as shown in Fig.~\ref{fig:current_fluctuations_vs_mu}(b), the LB current fluctuations $\langle \langle J^2 \rangle \rangle$ are different from those calculated by unravelling the non-additive master equations in the local or global bases,
$\langle \langle J^2 \rangle \rangle^{\text{c}}$ and $\langle \langle J^2 \rangle \rangle^{\text{d}}$. This supports our results demonstrating that these additive unravellings cannot be used to assess current fluctuations in the non-additive dynamics in the case of perfect detection. Apart from the bath's chemical potential, it is interesting to investigate the influence of other system's parameters, such as detuning or interdot tunnel coupling, on current fluctuations. This analysis is provided in Appendix~\ref{appsec:detuning}.

\begin{figure}[!t]
    \centering
    \includegraphics[width=1.0 \linewidth]{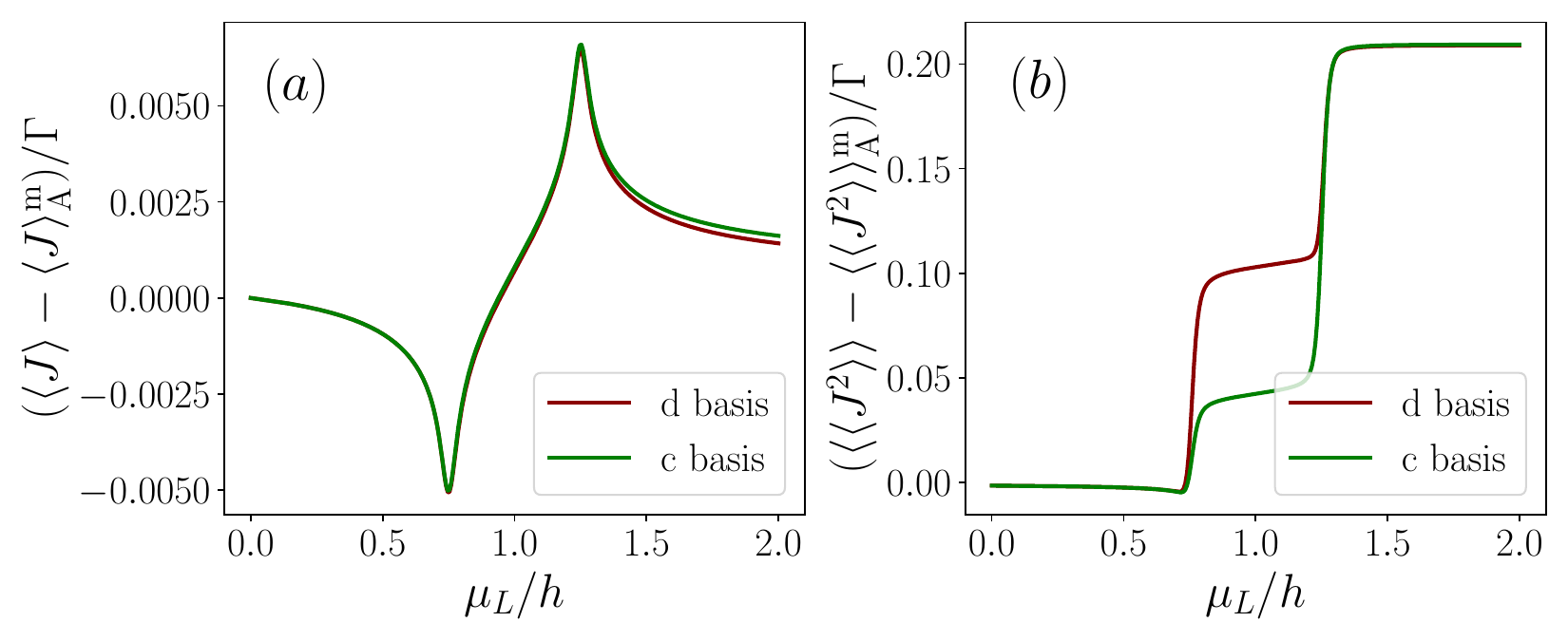} 
    \caption{Difference in steady-state current $(\langle J \rangle - \langle J \rangle^{\rm m}_{\rm A})/\Gamma$ (left) and its fluctuations $(\langle \langle J^2 \rangle \rangle - \langle \langle J^2 \rangle \rangle^{\rm m}_{\rm A})/\Gamma$ (right) in the non-additive picture  with the variable chemical potential $\mu_L$. The parameters are the same as in Fig.~\ref{fig:current_fluctuations_vs_mu}.}
    \label{fig:fom_coupling_gamma_vs_mu}
\end{figure}

Additionally, in order to investigate how non-additivity influences the average current and its fluctuations, we compare the LB values of these quantities with those obtained in the \emph{additive} counterpart of the non-additive dynamics. More specifically, we consider the total dissipator $\mathcal{L}$ without the cross-terms $\mathcal{L}_{LR}^{\rm m}$ in the local and global bases, respectively, for the evaluation of the steady state $\hat{\rho}_{\text{ss}}$. Once again, we use  Eqs.~\eqref{average_jump_current} and \eqref{current_fluctuations_eq_tot} to calculate these quantities in the corresponding additive dynamics ($\mathrm{A}$), which we denote by $\langle J \rangle_{\text{A}}^{\rm m}$ and $\langle \langle J^2 \rangle \rangle_\text{A}^{\rm m}$, where ${\text m} = {\text c},{\text d}$. In Fig.~\ref{fig:fom_coupling_gamma_vs_mu}, we plot the differences $\langle J \rangle - \langle J \rangle_\text{A}^{\rm m}$ and  $\langle \langle J^2 \rangle \rangle - \langle \langle J^2 \rangle \rangle_\text{A}^{\rm m}$ versus the chemical potential.

In Fig.~\ref{fig:fom_coupling_gamma_vs_mu}(a), we see that the mismatch of the average current for the additive counterpart with respect to the LB current is comparable for both the global basis  and local basis. This difference experiences a sharp rise, in both bases, when the chemical potential of the left reservoir lies between the eigenenergy levels $E_1,E_2$ of the system. On the other hand, as shown in Fig.~\ref{fig:fom_coupling_gamma_vs_mu}(b), the mismatch in the current fluctuations in different bases is different when $\mu_L/h$ is sandwiched between the system's eigenenergy levels, where the results are more sensitive to non-additivity. Both deviations of the current and its fluctuations are negative at small $\mu_L/h$ meaning that that these quantities are slightly larger when calculated for the both additive cases than in the non-additive dynamics. However, as $\mu_L/h$ increases, the differences become positive for large enough $\mu/h$. 

\section{Discussion and Concluding Remarks}
\label{sec:discussion}

In this work, we have demonstrated the breakdown of the quantum-jump approach for describing the fluctuations of the steady-state current in a non-additive open quantum system consisting of a two-site fermionic network in contact with two independent macroscopic thermal baths. Our main result is the impossibility of calculating current fluctuations by assuming additive quantum jump unravellings in the local or global basis in the case of perfect jump detection. To demonstrate this, we have shown that the evolution between jumps corresponds to a non completely-positive map whenever the master equation is non-additive. This implies that such ensemble dynamics, which describe the cooperative action of both reservoirs together, is incompatible with the trajectories that would arise from monitoring charges exchanged with each reservoir individually.

As a result of the breakdown of the quantum-jump approach, one must resort to alternative methods to calculate current fluctuations. To this end, we employ the Landauer-B\"uttiker formalism~\cite{datta2005quantum, di2008electrical, Buttiker_1986}. Assuming that the current fluctuations can be assessed with additive quantum jump unravellings, we have quantified the potential error for both observables in this case. While the difference between the current obtained using such unravellings and the Landauer-B\"uttiker benchmark is of order $1\%$, that for its fluctuations is of order $10 \%$. 

Another key finding is the calculation of fluctuations by applying additive unravellings of the non-additive master equation in the case of imperfect detection, that is, when some jumps go undetected.
In this scenario, one can accurately compute both the steady-state current and its fluctuations for certain parameter regimes in the site basis. In particular, at sufficiently high temperatures, the non-local term in the quantum master equation in the site basis can be neglected, resulting in the complete positivity of the no-click map, thereby ensuring the validity of additive unravellings for assessing fluctuations. 

Our results also provide an important experimental insight into current measurements in open quantum systems, such as quantum dot chains~\cite{Schuff_2026}. Measuring steady-state current fluctuations is possible by registering individual excitations emitted from a single dot to a detector, corresponding to an unravelling in the site basis. This is possible in quantum dots~\cite{Vigneau_2023} using local charge detectors, such as quantum point contacts~\cite{Elzerman_2003, DiCarlo_2004},  single-electron transistors~\cite{Schoelkopf_1998, Aassime_2001}, or quantum-dot charge sensors~\cite{Veldhorst_2015, House_2015}, by placing them near the measured quantum dot~\cite{Elzerman_2003, Field_1993}. Physically realising an unravelling in the energy eigenbasis would require time-resolved calorimetry~\cite{Gasparinetti2015,Berg2015, Brange2018, Karimi2020} of emitted electrons, which is far more technically demanding. Our Fig.~\ref{fig:cpness_of_imd} corresponds to the local measurement scheme.

Overall, our work provides a pathway toward understanding how to measure steady-state current fluctuations in open quantum systems. Since a particular additive unravelling corresponds to a specific measurement scheme, it remains unclear whether unravelling the non-additive master equation in some other basis or manner, for example, via a non-Markovian unravelling~\cite{Breuer_2004, Piilo_2008, Settimo_2026}, a diffusive unravelling~\cite{Barchielli_2010}, a partially observed unravelling when the detector only accesses a coarse-grained charge signal, or via an unravelling with correlated quantum jumps~\cite{Salatino_2026}  -- and thus performing the associated measurement, for instance, via rf reflectometry~\cite{Vigneau_2023}  -- would enable the exact determination of the current fluctuations when some jumps go undetected. In this regard, it is still an open question whether one can compute steady-state current fluctuations using the quantum jump approach under imperfect detection and perform their experimental measurement.

\section*{Acknowledgments}
I. K. thank Amr Dodin, Giuliano Benenti, Kacper Prech, Hari K. Yadalam, Farhan T. Chowdhury, and Hayden Zammit for fruitful discussions. M. T. M. is supported by a Royal Society University Research Fellowship. This project is co-funded by the European Union (Quantum Flagship project ASPECTS, Grant Agreement No.\ 101080167) and UK Research and Innovation (UKRI). Views and opinions expressed are however those of the authors only and do not necessarily reflect those of the European Union, Research Executive Agency or UKRI. Neither the European Union nor UKRI can be held responsible for them.

\section*{Data Availability}

All the numerical results presented in this work were obtained using the Python framework QuTiP~\cite{Johansson_2012, Johansson_2013, Lambert_2026}, which are available on GitHub~\cite{myGit2026}.

\bibliographystyle{apsrev4-1fixed_with_article_titles_full_names.bst}

\bibliography{bibbib}
\onecolumngrid
\newpage
\appendix
\section{Derivation of the non-additive quantum master equation}
\label{appsec:derivation}
In this Appendix we provide the main steps leading to the non-additive quantum master equations in Eqs.~\eqref{eq:decompositions} and \eqref{eq:decompositions_d} as derived in Ref.~\cite{Mitchison_2018}, to which we refer the reader for further details.
\subsection{Reservoirs Description}
\label{appsec:Reservoirs}

\begin{figure}[h!]
    \centering
    \includegraphics[width=1.0\linewidth]{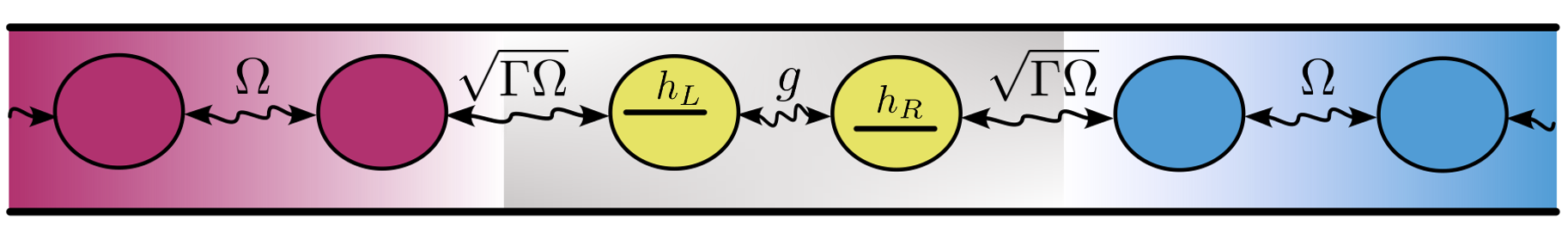}
    \caption{Model of reservoirs as 1D tight-binding chains, where $\Omega$ and $\sqrt{\Gamma\Omega}$ are the intra-reservoir and system-reservoir tunnelling energies, respectively.}
    \label{fig:reservoirs_illustration}
\end{figure}

Each bath in our model is a particle reservoir with the Hamiltonian $\hat{H}_\alpha$ such that $\hat{H}_B =\hat{H}_L+\hat{H}_R$. Both Hamiltonians $\hat{H}_\alpha$ describe a 1D fermionic tight-binding model on $M$ sites with hopping amplitude $\Omega >0$, also known as a uniform chain, which is schematically shown in Fig.~\ref{fig:reservoirs_illustration}.
The explicit forms of the Hamiltonians $\hat{H}_\alpha$ are
\begin{eqnarray}
    \hat{H}_\alpha =  -\frac{\Omega}{2}\sum_{m =1}^{M-1} (\hat{A}_{m, \alpha}^{\dagger} \hat{A}_{m+1, \alpha} + \hat{A}_{m+1, \alpha}^{\dagger} \hat{A}_{m, \alpha}) =  \sum_q \omega_q \hat{a}^\dagger_{q, \alpha} \hat{a}_{q, \alpha},
\label{eq:bath_ham}
\end{eqnarray}
where $\hat{A}_{m, \alpha}$ annihilates a fermion localised on site $m$ of reservoir $\alpha$ and satisfies \( \{\hat{A}_{m, \alpha}, \hat{A}_{m', \alpha'}^\dagger\} = \delta_{mm'} \delta_{\alpha \alpha'} \) and \( \{\hat{A}_{m, \alpha}, \hat{A}_{m', \alpha'}\} =0 \). The canonical transformation $\hat{a}_{q,\alpha}$
\begin{equation}
    \hat{a}_{q,\alpha} = \sqrt{\frac{2}{M+1}} \sum_{m=1}^M \sin{(qm)} \hat{A}_{m,\alpha}
\label{eq:transform}
\end{equation}
leads to the diagonal form of the Hamiltonian. Here, the ladder operators $\hat{a}_{q, \alpha}$ corresponds to fermionic modes characterised by a dimensionless wave number $q = \pi k/(M+1)$, $k=1, \ldots M$, with dispersion relation $\omega_q = -\Omega \cos(q)$.

The interaction between the system and reservoir is carried out via tunnelling of fermions between the terminal site of each bath and the neighbouring site of system $S$. For convenience, let us define the tunnelling energy as $\sqrt{\Gamma \Omega}$, with $\Gamma$ setting the overall frequency scale of the dissipative dynamics. This allows us to write down the explicit form of the interaction Hamiltonian 
\begin{eqnarray}
    \hat{H}_{SB} =  \frac{\sqrt{\Gamma\Omega}}{2}\sum_{\alpha =L,R}(\hat{c}_{\alpha}^{\dagger} \hat{A}_{1, \alpha} + \hat{A}_{1, \alpha}^{\dagger} \hat{c}_{\alpha}) = 
    \sum_{\alpha = L,R} \sum_q (\tau_q \hat{c}^\dagger_{\alpha} \hat{a}_{q, \alpha} +\tau_q^* \hat{a}_{q, \alpha}^\dagger \hat{c}_{\alpha}),
\label{eq:int_ham}
\end{eqnarray}
where $\tau_q = \sqrt{\Gamma\Omega/(M+2)} \sin{(q)}$ is the tunnelling coupling.

Investigating the system in a basis that diagonalises $\hat{H}_S$, given by Eq.~\eqref{eq:system_ham}, is simpler than in the site basis. Following this, we arrange the ladder
operators as follows, $\hat{\mathbf{c}}= (\hat{c}_L, \hat{c}_R)^{\text{T}}$ and $\hat{\mathbf{a}}_q= (\hat{a}_{q,L}, \hat{a}_{q,R})^{\text{T}}$, and introduce a new canonical set of ladder operators, namely, $\hat{\mathbf{d}}= (\hat{d}_1, \hat{d}_2)^{\text{T}} = \text{R} \hat{\mathbf{c}}$ and $\hat{\mathbf{b}}_q= (\hat{b}_{q,1}, \hat{b}_{q,2})^{\text{T}} = \text{R} \hat{\mathbf{a}}_q$, connected via the following orthogonal rotation
\begin{equation}
\mathrm{R} = \frac{1}{\sqrt{2\Delta}}
\left(
\begin{array}{cc}
\sqrt{\Delta + \delta} & -\sqrt{\Delta - \delta} \\ \sqrt{\Delta - \delta} & \sqrt{\Delta + \delta}
\end{array}
\right),
\label{eq:rotation_matrix}
\end{equation}
with $\Delta=\sqrt{\delta^2+g^2}$. Such a transformation divides the Hamiltonian into two independent parts $\hat{H}=\hat{H}_1+\hat{H}_2$, in which
\begin{equation}
\hat{H}_j = E_j \hat{d}_j^\dagger \hat{d}_j + \sum_q \Big( \omega_q \hat{b}_{q,j}^\dagger \hat{b}_{q,j} + \tau_q \hat{d}_j^\dagger \hat{b}_{q,j} + \tau_q^* \hat{b}_{q,j}^\dagger \hat{d}_j \Big),
\label{eq:new_part}
\end{equation}
where $E_j = h \pm \frac{\Delta}{2}$ are the single-particle energy eigenvalues of $\hat{H}_S$. For simplicity, let us assume $E_j > 0$ which means that $h_\alpha > 0$ and $g < \sqrt{h_L h_R}$. 

The influence of the reservoirs on the system is determined by the spectral density of the thermal baths
\begin{equation}
\mathcal{S}(\omega) = \sum_q |\tau_q|^2 \delta(\omega - \omega_q).
\label{eq:spectral_function}
\end{equation}
When $M \rightarrow \infty$, the spacing between adjacent wave vectors $\Delta q = \pi/(M+1)$ tends to zero that results in $q$ being a continuous variable with values in the first Brillouin zone, $ q \in [0, \pi]$, which enables us to replace $\sum_q \Delta q$ by $\int \mathrm{d}q$ and transform $\mathcal{S}(\omega)$ to 
\begin{equation}
\mathcal{S}_{\text{N}}(\omega) = \frac{\Gamma}{2 \pi} \sqrt{1 - \frac{\omega^2}{\Omega^2}} \Theta(\Omega - |\omega|),
\label{eq:spectral_function_full}
\end{equation}
where $\Theta(x)$ is the Heaviside unit step function, and the subscript $\text{N}$ denotes the Newns spectral density, introduced in Ref.~\cite{Newns_1969}. Such a form of the spectral density sets a spectral bandwidth $\Omega$ bringing about vacuum correlation of time order $\Omega^{-1}$.

\subsection{Structure of the Dissipator in the Exact Master Equation}
\label{appsec:dissipator}

As we discussed in the main text, the dynamics of the open quantum system under consideration are governed by the exact master equation in Eq.~\eqref{eq:exact_master_equation}, with the steady-state dissipator $\mathcal{L}$ of the following form 
\begin{equation}
\mathcal{L}\hat{\rho}_S =\sum_{j, k =1}^2 \Lambda^-_{jk} \left(\hat{d}_k^\dagger \hat{\rho}_S \hat{d}_j - \frac{1}{2} \{\hat{d}_j \hat{d}_k^\dagger, \hat{\rho}_S\} \right) + \sum_{j, k =1}^2 \Lambda^+_{jk} \left(\hat{d}_j \hat{\rho}_S \hat{d}_k^\dagger - \frac{1}{2} \{\hat{d}_k^\dagger \hat{d}_j, \hat{\rho}_S\}\right). 
\label{eq:dissipator}
\end{equation}
The rate matrices $\Lambda^{\pm}$ have the components
\begin{equation}
\Lambda_{jk}^{+} = \mathrm{i} \int \mathrm{d}\omega \, \mathcal{S}_N(\omega) F_{jk}(\omega) 
\left[ \frac{1}{E_j - \omega + \mathrm{i} \Gamma_j /2} - \frac{1}{E_k - \omega - \mathrm{i} \Gamma_k /2} \right],
\label{eq:rate_matrices}
\end{equation}
and $\Lambda_{ij}^{-}= \Gamma_{j} - \Lambda_{ij}^{+}$ with $\text{F} = \text{R}~\text{diag}[f_L(\omega),f_R(\omega)]~\text{R}^\text{T}$.

For clarity, let us factorise the initial conditions in the form $\hat{\rho}(0)= \hat{\rho}_S(0)\otimes\hat{\rho}_L\otimes \hat{\rho}_R$ and initialise reservoir $\alpha$ in the Gibbs state. In this case, the system will relax to a unique steady state if $\sqrt{E_j^2+ \Gamma \Omega} < \Omega$. 
Throughout the rest of the paper, we assume the Newns spectral density~Eq.~(\ref{eq:spectral_function_full}) and $\Gamma \ll \Omega$, the so-called  exponential-propagator approximation (EPA).

\subsubsection{Energy Eigenbasis}\label{Appendix_B1}
Since we deal with the asymptotic dynamics, we refer to the dissipator $\mathcal{L}$ as asymptotic, which can be decomposed as
\begin{equation}
\mathcal{L} = \mathcal{L}_L^{\rm d}+ \mathcal{L}_R^{\rm d}+\mathcal{L}_{LR}^{\rm d}. 
\label{eq:assimptotic_dissipator}
\end{equation}

Each of these Lindblad generators $\mathcal{L}_\alpha^{\rm d}$ has the form
\begin{equation}
\mathcal{L}_{\alpha}^{\rm d} = \sum_{j=1}^{2} \left( \gamma_{\alpha,j}^{-} \mathcal{D}[\hat{d}_j] + \gamma_{\alpha,j}^{+} \mathcal{D}[\hat{d}_j^{\dagger}] \right),
\label{eq:generators}
\end{equation}
where $\mathcal{D}[\hat{d}]~\bullet = \hat{d}\bullet\hat{d}^{\dagger} - \frac{1}{2}\{ \hat{d}^{\dagger}\hat{d},\bullet\}$ is a Lindblad dissipator and the decay and gain rates are given by 
\begin{eqnarray}
\gamma_{\alpha, j}^{-}=\Gamma_{\alpha, j} \int \mathrm{~d} \omega \ \varphi_j(\omega)[1-f_{\alpha}(\omega)] \\ \nonumber
\gamma_{\alpha, j}^{+}=\Gamma_{\alpha, j} \int \mathrm{~d} \omega \ \varphi_j(\omega)f_{\alpha}(\omega),
\label{eq:rates}
\end{eqnarray}
with $\Gamma_{L,j}= \Gamma_{j}^2R_{1j}^2$, $ \Gamma_{R,j}= \Gamma_{j}^2R_{2j}^2$, and $R_{ij}$ being the matrix element of the matrix $\mathrm{R}$. Here, $\varphi_j(\omega)$ is the probability distribution of excitation energies associated with the state $\hat{d}^\dagger_j \ket{0}$ and $\ket{0}$ is the vacuum state, i.e. $\hat{N} \ket{0}=0$, where $\hat{N}$ is the total number of fermions in the system. These functions have the form 
\begin{eqnarray}
\varphi_j(\omega)=\frac{1}{1-\Gamma / \Omega} \frac{\mathcal{S}_{\mathrm{N}}(\omega)}{\left(\omega-E_j^{\prime}\right)^2+\Gamma_j^2 / 4} \approx
\frac{\mathcal{S}_{\mathrm{N}}(\omega)}{\left(\omega-E_j\right)^2+\Gamma_j^2 / 4},
\label{eq:propability_destribution}    
\end{eqnarray}
where the approximate equality holds under the EPA; $E_j^{\prime} \approx E_j$ are the shifted energies, and $\Gamma_j = 2 \pi \mathcal{S}_{N}(E_j)$ are the decay rates.

The interference contribution has the following form
\begin{equation}
\mathcal{L}_{LR}^{\rm d} \hat{\rho}_S =\sum_{j \neq k} \Lambda^+_{jk} \left[ \left(\hat{d}_k^\dagger \hat{\rho}_S \hat{d}_j - \frac{1}{2} \{\hat{d}_j \hat{d}_k^\dagger, \hat{\rho}_S\} \right) - \left(\hat{d}_j \hat{\rho}_S \hat{d}_k^\dagger - \frac{1}{2} \{\hat{d}_k^\dagger \hat{d}_j, \hat{\rho}_S\}\right) \right],
\label{eq:interference}
\end{equation}
where ${\Lambda}_{12}^{+}=({\Lambda}_{21}^{+})^* = \xi + \mathrm{i}\eta$ with
\begin{eqnarray}
\xi = \frac{g}{4\Delta} \int \mathrm{d}\omega \sum_{j=1}^{2} \Gamma_j \varphi_j(\omega) [f_L(\omega) - f_R(\omega)], \\
\eta = \frac{g}{2\Delta} \int \mathrm{d}\omega [(E_1 - \omega) \varphi_1(\omega) - (E_2 - \omega) \varphi_2(\omega)] [f_L(\omega) - f_R(\omega)].
\label{eq:interference_re_im_parts}
\end{eqnarray}

Note that $\mathcal{L}_{LR}^{\rm d}$ provides no positive evolution and is proportional to the difference in distribution functions $f_\alpha(\omega)$. This means that we can neglect this term only if $f_L(\omega) \approx f_R(\omega)$ when $\varphi_j(\omega)$ is sufficiently far from zero. The standard picture of the global GKSL master equation disregards this term because of the secular approximation; however, $\mathcal{L}_{LR}^{\rm d}$ has comparable values with $\mathcal{L}_\alpha^{\rm d}$ when the secular approximation holds, for example, in the quantum-optical limit, i.e. $\Gamma \ll \Delta$. 

\subsubsection{Site Eigenbasis}
The asymptotic dissipator $\mathcal{L}$ can be studied in a local picture, in which it can be decomposed in the site basis $\hat{c}_\alpha$ as 
\begin{equation}
\mathcal{L} = \mathcal{L}_L^{\rm c}+ \mathcal{L}_R^{\rm c}+\mathcal{L}_{LR}^{\rm c},
\label{eq:local_decomposition}  
\end{equation}
where the local dissipators are represented by Lindblad generators acting on a single site $\alpha$ in a form of 
\begin{equation}
 \mathcal{L}_\alpha^{\rm c} = \bar{\gamma}_\alpha^- \mathcal{D}[\hat{c}_\alpha] + \bar{\gamma}_\alpha^+ \mathcal{D}[\hat{c}_\alpha^\dagger],
\label{eq:local_generators}  
\end{equation}
with $\overline{\gamma}^{\pm}_{\alpha} = {\gamma}^{\pm}_{\alpha, 1} + {\gamma}^{\pm}_{\alpha, 2}$. The contribution $\mathcal{L}_{LR}^{\rm c}$ corresponds to delocalised processes involving both sites
\begin{equation}
\mathcal{L}_{LR}^{\rm c} \hat{\rho}_S = \sum_{\alpha \neq \alpha'} \left[ \bar{\Lambda}_{\alpha \alpha'}^+ \left( \hat{c}_{\alpha'}^\dagger  \hat{\rho}_S \hat{c}_\alpha - \frac{1}{2} \{ \hat{c}_\alpha \hat{c}_{\alpha'}^\dagger, \hat{\rho }_S \} \right) + \bar{\Lambda}_{\alpha \alpha'}^- \left( \hat{c}_{\alpha} \hat{\rho }_S \hat{c}_{\alpha'}^\dagger - \frac{1}{2} \{ \hat{c}_{\alpha'}^\dagger \hat{c}_\alpha, \hat{\rho }_S \} \right) \right],
\label{eq:delocal_dissipator}  
\end{equation}    
where ${\bar{\Lambda}}_{LR}^{\pm}=({\bar{\Lambda}}_{RL}^{\pm})^*$ and
\begin{align}
\text{Re} [\bar{\Lambda}^{+}_{LR}] &= \frac{g}{4\Delta} \int d\omega \left[ \Gamma_2 \varphi_2(\omega) - \Gamma_1 \varphi_1(\omega) \right] \left[ f_L(\omega) + f_R(\omega) \right], \\
\text{Im} [\bar{\Lambda}^{+}_{LR}] &= \frac{g}{2\Delta} \int d\omega \left[ (E_1 - \omega) \varphi_1(\omega) - (E_2 - \omega) \varphi_2(\omega) \right] \left[ f_L(\omega) - f_R(\omega) \right], \\
\text{Re} [\bar{\Lambda}^{-}_{LR}] &= \frac{g}{4\Delta} \int d\omega \left[ \Gamma_2 \varphi_2(\omega) - \Gamma_1 \varphi_1(\omega) \right] \left[ 2 - f_L(\omega) - f_R(\omega) \right], \\
\text{Im} [\bar{\Lambda}^{-}_{LR}] &= -\text{Im} [\bar{\Lambda}^{+}_{LR}].
\label{eq:real_im_parts}
\end{align}

The cross-term $\mathcal{L}_{LR}^{\rm c}$ is related to the difference of the probability distributions of excitation energies $\varphi_j(\omega)$. That is said, $\mathcal{L}_{LR}^{\rm c}$ shows how the occupation numbers $f_\alpha(\omega)$ of reservoir states are different between the separate energies ranges corresponding to the two distributions $\varphi_j(\omega)$. One can neglect $\mathcal{L}_{LR}^{\rm c}$ only in the white-noise limit, in which $\Delta \ll \Omega$ and the distribution functions $f_\alpha(\omega)$ are constant over the energy range where $\varphi_j(\omega)$ are non-zero.

\section{Vectorisation}
\label{appsec:Vectorisation}

Some of our results involve the vectorised notation of operators. For this purpose, we explain the details of this procedure in the current appendix. To vectorise operators, we will transform all of them into vectors in Liouville space, stacking their columns in the following way
\begin{align}
    \operatorname{vec}\left(\begin{array}{ll}a & b \\ c & d\end{array}\right)=\left(\begin{array}{l}a \\ c \\ b \\ d\end{array}\right)
\end{align}
using the notation
\begin{align}
    |\hat{\rho}\rangle\!\rangle=\operatorname{vec}(\hat{\rho}),
\label{eq:vector_rho}
\end{align}
and all superoperators into operators using the convention 
\begin{align}
    \operatorname{vec}(A \hat{\rho} C)=\left(C^{\mathrm{T}} \otimes A\right) \operatorname{vec}(\hat{\rho})
\label{eq:vectorisation_rule}
\end{align}
The time-dependent state can be written as
\begin{align}
    |\hat{\rho}(t)\rangle\!\rangle =|\hat{\rho}_{\mathrm{ss}}\rangle\!\rangle + \sum_{i>0}e^{\lambda_i t}|x_i\rangle\!\rangle \langle\!\langle y_i|\hat{\rho}_0\rangle\!\rangle,
\label{eq:density_matrix}
\end{align}
where $|x_i\rangle\!\rangle$ and $ \langle\!\langle y_i|$ are the right and left eigenvectors of the Liouvillian corresponding to eigenvalue $\lambda_i$ (with $\operatorname{Re}(\lambda_i)< 0$), fulfilling $\langle\!\langle y_i|x_i\rangle\!\rangle=\delta_{ij}$\footnote{Note that these eigenvectors are not complex conjugations of each other as the Liouvillian is non-hermitian, so $|x_i\rangle\!\rangle^\dagger\neq\langle\!\langle y_i|$.}, and $|\hat{\rho}_0\rangle\!\rangle$ is the initial state. 

The average current corresponding to the current superoperator $\mathcal{J}$ can be expressed in Liouville space via 
\begin{align}
    J(t)=\langle\!\langle \mathbf{1}|\mathcal{J}|\hat{\rho}(t)\rangle\!\rangle,
\label{eq:general_current}
\end{align}
where $\langle\!\langle \mathbf{1}|$ is the vectorised identity operator. For jump processes, as we have them here, the variance (or noise) can be written in vectorised notation as~\cite{Landi_2024}
\begin{align}
    \langle\!\langle J^2\rangle\!\rangle = K - 2 \langle\!\langle \mathbf{1}| \mathcal{J} \mathcal{L}^+ \mathcal{J} |\hat\rho_{\mathrm{ss}}\rangle\!\rangle
\label{eq:general_flucutations}
\end{align}
where $\mathcal{L}^+=\sum_{i>0}\frac{1}{\lambda_i}|x_i\rangle\!\rangle \langle\!\langle y_i|$ is the Drazin pseudoinverse. 

\section{Investigation of Complete Positivity of No-jump Quantum Map in the perfect detection scenario}\label{appsec:no_jump}

\begin{figure}[h]
    \centering
    \includegraphics[width=0.7\linewidth]{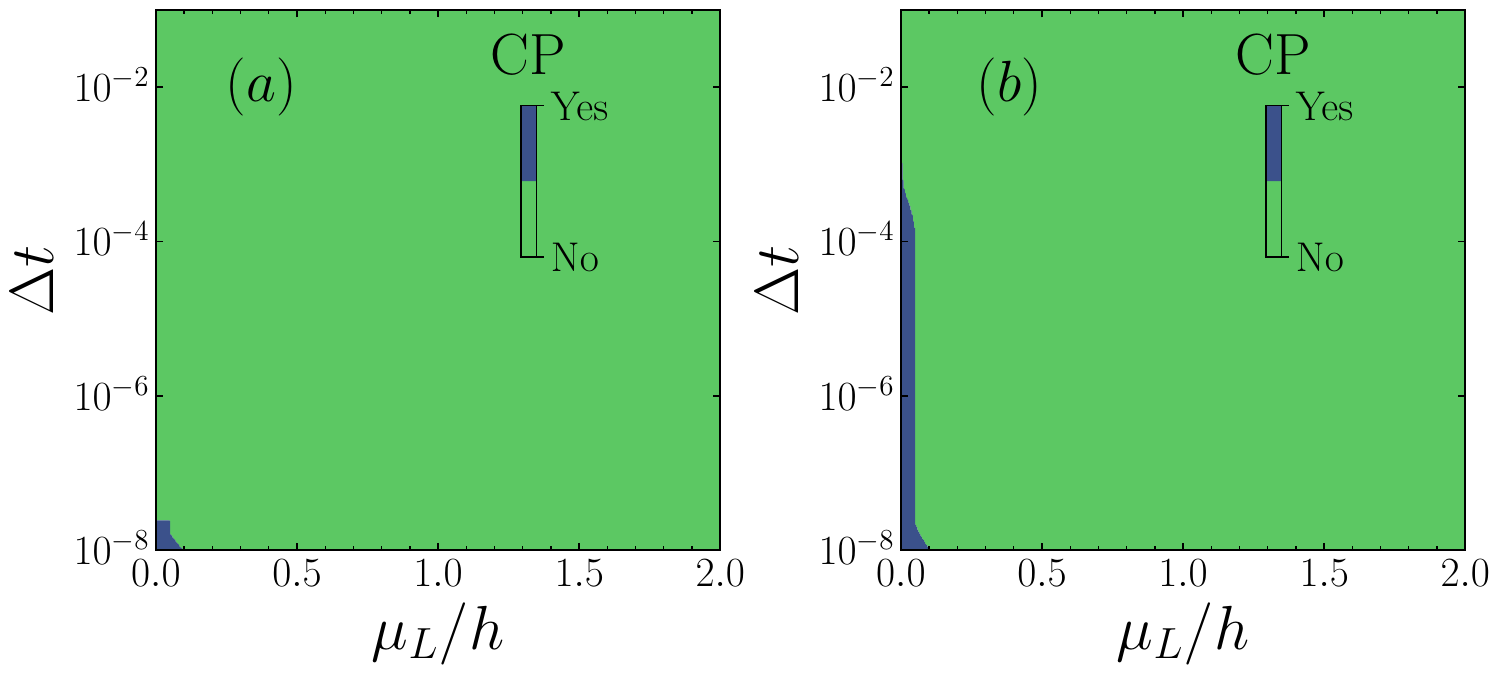}
    \caption{Complete positivity (CP) of the quantum maps $\mathcal{M}_0^{\rm c}$ and $\mathcal{M}_0^{\rm d}$ as a function of the chemical potential $\mu_L/h$ and time between jumps $\Delta t$, depicted in a logarithmic scale. Parameters: $h=1$, $g=0.5$, $\delta =0.0$, $\Delta= \sqrt{\delta^2+g^2}$, $\Gamma = 0.02 \Delta$, $\Omega = 10$, $T_L=T_R = 0.02 \Delta$, $\mu_R=0$. The blue (green) colour denotes that the Choi matrix of the quantum map has non-negative (negative) eigenvalues at the given time $\Delta t$.
    } 
    \label{fig:cpness_of_jo}
\end{figure}

In this appendix, we present the results of the numerical test for complete positivity of the no-jump maps $\mathcal{M}_0^{\rm c}$ and $\mathcal{M}_0^{\rm d}$, which are displayed in Fig.~\ref{fig:cpness_of_jo} as a function of $\mu_L/h$ and $\Delta t$. 

As the time between consecutive jumps $\Delta t$ is a stochastic variable that can take arbitrary values, we conclude that neither $\mathcal{M}_0^{\rm c}$ nor $\mathcal{M}_0^{\rm d}$ correspond to a CP evolution between jumps. This implies that perfectly detecting jumps as described by the additive unravellings considered here, while preserving the non-additive dynamics generated by the dissipators in Eqs.~\eqref{eq:interference_local} and~\eqref{eq:interference_global} leads to unphysical processes.

\section{Influence of Local Energies' Detuning and Interdot Coupling on Average Steady-State Current and Fluctuations}
\label{appsec:detuning}

\begin{figure}[h]
    \centering
    \includegraphics[width=\textwidth]{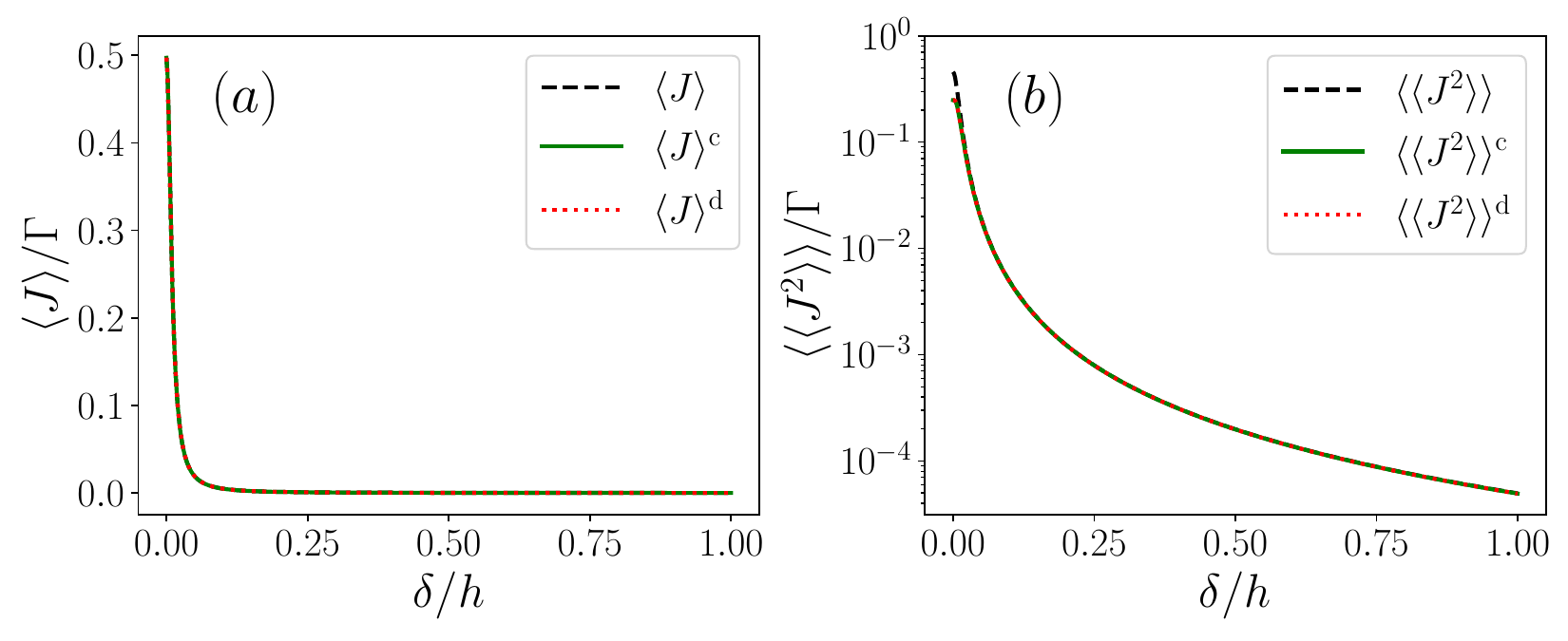}
    \caption{Average steady-state current $\langle J \rangle/\Gamma$ (a) and its fluctuations $\langle \langle J^2 \rangle \rangle/\Gamma$ (b) through the double quantum dot in the limit $\Gamma \ll \Delta \ll \Omega$ with the variable local energy levels detuning $\delta$. The fluctuations are depicted in a logarithmic scale for better visualisation. Parameters: $h=1$, $g =0.01$, $\Gamma = 0.02 \Delta$, $\Omega = 10$, $T_L=T_R =0.02 \Delta$, $\mu_L=2h$, $\mu_R=0$.}
    \label{fig:current_fluctuations_vs_detuning}
\end{figure}

As we discussed in the main text, the energy window set by the chemical potentials of the reservoirs has a dramatic impact on the fluctuations, resulting in a strong disagreement between the QMEs and LB approaches when the chemical potential of one bath crosses the energy levels of the system assuming the other chemical potential to be fixed. Now, we illustrate the role of the system's parameters, such as the interdot coupling and the detuning between the quantum-dot energy levels, on the steady-state current and its fluctuations, computed via the Landauer-B\"uttiker approach, the master equation in the local and global bases, which are shown in Fig.~\ref{fig:current_fluctuations_vs_detuning}.

It is clearly seen that the detuning $\delta$ has no effect on the current (Fig.~\ref{fig:current_fluctuations_vs_detuning} (a)), while it leads to a slight difference in fluctuations for small values of $\delta$, as shown in Fig.~\ref{fig:current_fluctuations_vs_detuning} (b). For small values of $\delta$, $\mathcal{L}_{LR}^{\rm m} \sim g/\sqrt{\delta^2+g^2}$ is not negligible, so the master equation~\eqref{eq:exact_master_equation} is a GKSL equation but not purely additive. By increasing $\delta$, one can neglect $\mathcal{L}_{LR}^{\rm m} \approx 0$, which brings the total dissipator $\mathcal{L}$ to its additive Lindblad form. Both the interdot coupling $g$ and  energy levels detuning $\delta$ are reasonable to scale with respect to the energy levels' mean $h$, since $\delta$ characterises how the dots' energies deviate from one another relative to this reference and $g$ describes how strongly the dots are coupled with each other.

\begin{figure}[h!]
    \centering
    \includegraphics[width=\textwidth]{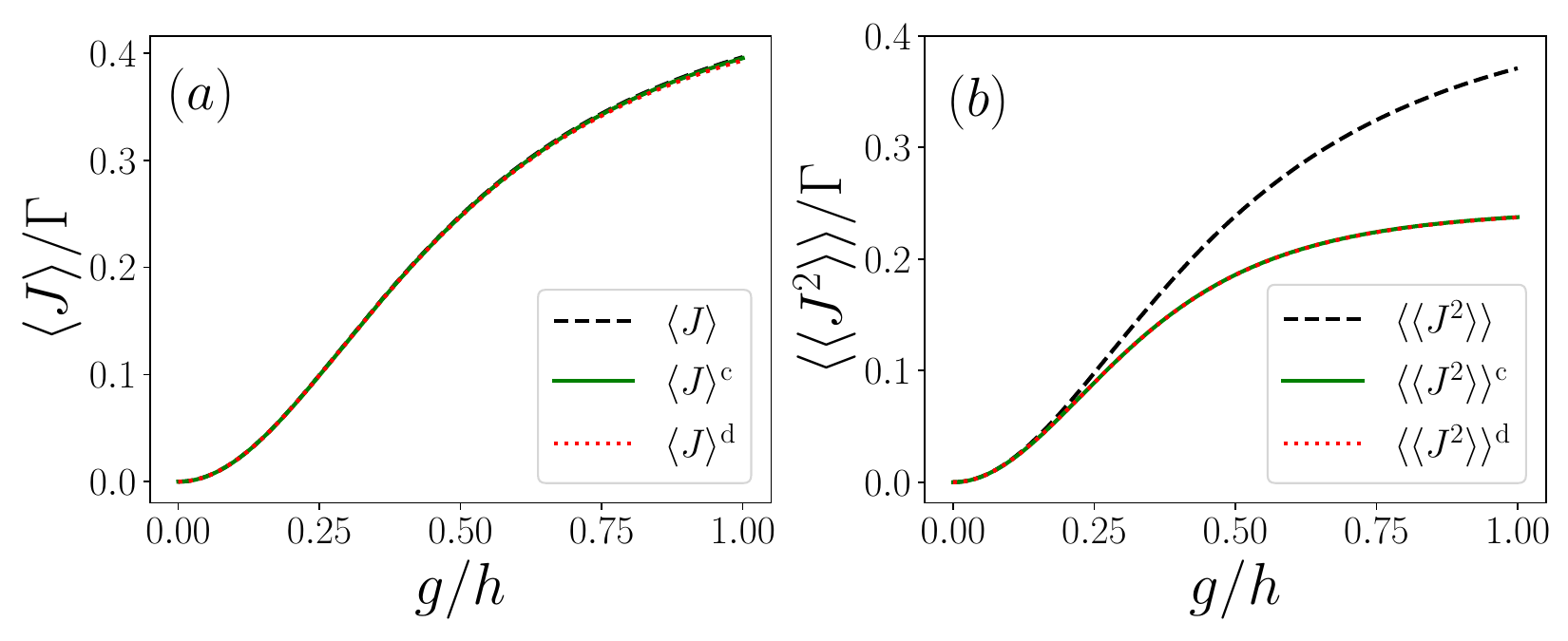}
    \caption{Average steady-state current $\langle J \rangle/\Gamma$ (a) and its fluctuations $\langle \langle J^2 \rangle \rangle/\Gamma$ (b) through the double quantum dot in the limit $\Gamma \ll \Delta \ll \Omega$ with the variable tunnel coupling $g$ between the quantum dots. The subscripts $\text{c}$ and $\text{d}$ refer to the site and energy eigenbasis, respectively. Parameters: $h=1$, $\delta =0.5$, $\Gamma = 0.02 \Delta$ and $\Omega = 10$, $T_L=T_R =0.02 \Delta$, $\mu_L=2h$, $\mu_R=0$.}
    \label{fig:current_fluctuations_vs_g}
\end{figure}

Another system parameter whose role is interesting to examine is the interdot tunnel coupling. When this coupling is weak, $g \ll h$, the system behaves as two separate dots since the interaction between them is weak. This enable us to neglect $\mathcal{L}_{LR}^{\rm m} \approx 0$ and apply the quantum jump approach. Fig.~\ref{fig:current_fluctuations_vs_g} depicts the average current and its fluctuations as a function of the scaled coupling $g/h$ using the Landauer-B\"uttiker approach, the master equation in the local and global bases.
While in the quantum-optical limit, at low temperatures, the average current obtained from the master equation in both bases demonstrates nearly perfect agreement with its LB benchmark (Fig.~\ref{fig:current_fluctuations_vs_g} (a)), the fluctuations calculated in the LB approach partially agree with those from the master equation for weak coupling $g/h \ll 1$, with a significant discrepancy emerging between two approaches at strong coupling (Fig.~\ref{fig:current_fluctuations_vs_g} (b)). In this regime, the results obtained from both bases are nearly identical.

\section{Analysis of Complete Positivity of Quantum Map in the imperfect detection scenario}
\label{appsec:positivity_complete_positivity}

In this section, we investigate the complete positivity of quantum maps that govern a quantum jump evolution in the case of imperfect detection. Suppose we have an unravelling with four jump operators in the site basis: $\hat{L}_1^1 = \sqrt{\bar{\gamma}^{-}_L\eta_1^{\rm c}} \hat{c}_L$, $\hat{L}_2^1 = \sqrt{\bar{\gamma}^{+}_L\eta_2^{\rm c}} \hat{c}_L^\dagger$, $\hat{L}_3^1 = \sqrt{\bar{\gamma}^{-}_R\eta_3^{\rm c}} \hat{c}_R$, and $\hat{L}_4^1 = \sqrt{\bar{\gamma}^{+}_R\eta_4^{\rm c}} \hat{c}_R^\dagger$, where each jump can be measured with an efficiency $\eta_k^{\rm c} \in [0,1]$ and $\eta_k^{\rm c} = \tilde{\gamma}^{\pm}_\alpha/\bar{\gamma}^{\pm}_\alpha$, i.e. $\eta_k^{\rm c} = 1$ corresponds to the perfect efficiency~\cite{Landi_2024}. Using this unravelling, we can write the system of equations for the average steady-state current and its fluctuations $\langle J \rangle_\alpha$ and $\langle \langle J^2 \rangle \rangle_\alpha$ as
\begin{eqnarray}
\label{eq:cf}
\begin{cases}
\langle J \rangle_\alpha = \tilde{\gamma}^{+}_{\alpha} - (\tilde{\gamma}^{+}_{\alpha} + \tilde{\gamma}^{-}_{\alpha}) \langle \hat{c}_\alpha^\dagger\hat{c}_\alpha \hat{\rho}_{\mathrm{ss}} \rangle\\
\langle \langle J^2 \rangle \rangle_\alpha =  K_\alpha - 2 \langle \langle \mathbf{1} | \mathcal{J} \mathcal{L}^{+}\mathcal{J} | \hat{\rho}_{\mathrm{ss}} \rangle \rangle,
\end{cases}
\end{eqnarray}
where $\mathcal{J} \hat{\rho}_{\mathrm{ss}} = \sum_{k} \nu_k \hat{L}_k^{\dagger} \hat{L}_k \hat{\rho}_{\mathrm{ss}}$ is the current operator, $K_\alpha = \sum_{k} \nu_k^2 \mathrm{Tr} \{ \hat{L}_k^{\dagger} \hat{L}_k \hat{\rho}_{\mathrm{ss}} \} = \tilde{\gamma}^{+}_{\alpha} - (\tilde{\gamma}^{+}_{\alpha} - \tilde{\gamma}^{-}_{\alpha}) \langle \hat{n}_\alpha \rangle_{ss}$ is the dynamical activity, and the weights $\nu_k$ are equal to $+1$ if we create a fermion in the system and to $-1$ when we destroy a fermion in the system. Note that $\langle J \rangle_L = - \langle J \rangle_R$ and $\langle \langle J^2  \rangle \rangle_L = \langle\langle J^2 \rangle\rangle_R$.   

Substituting the expression for the current operator into the second equation of the system~\eqref{eq:cf} and using the fermionic anticommutation relation, we obtain the following expression 
\begin{eqnarray}
\begin{cases}
\langle J \rangle_\alpha = \tilde{\gamma}^{+}_{\alpha} - (\tilde{\gamma}^{+}_{\alpha} + \tilde{\gamma}^{-}_{\alpha}) \langle \hat{n}_\alpha \rangle_{ss}\\
\langle \langle J^2 \rangle \rangle_\alpha = \tilde{\gamma}^{+}_{\alpha} - (\tilde{\gamma}^{+}_{\alpha} - \tilde{\gamma}^{-}_{\alpha}) \langle \hat{n}_\alpha \rangle_{ss} - \zeta~ (\tilde{\gamma}^{+}_{\alpha})^2 + (\beta + \xi)~ \tilde{\gamma}^{+}_{\alpha}\tilde{\gamma}^{-}_{\alpha} - \delta(\tilde{\gamma}^{-}_{\alpha})^2,  
\label{eq:final}    
\end{cases}
\end{eqnarray}
where $\langle \hat{n}_\alpha \rangle_{ss} = \mathrm{Tr}  \{ \hat{c}_\alpha^\dagger\hat{c}_\alpha \hat{\rho}_{\mathrm{ss}} \}$, $\zeta = \mathrm{Tr} \left \{\hat{c}_\alpha^\dagger \left [ \mathcal{L}^+ \{ \hat{c}_\alpha^\dagger \hat{\rho}_{\mathrm{ss}} \hat{c}_\alpha\} \right] \hat{c}_\alpha \right \}$, $\beta = \mathrm{Tr} \left \{  \hat{c}_\alpha^\dagger \left [ \mathcal{L}^+ \{ \hat{c}_\alpha \hat{\rho}_{\mathrm{ss}} \hat{c}_\alpha^\dagger\} \right] \hat{c}_\alpha \right \}$, $\xi = \mathrm{Tr} \left \{  \hat{c}_\alpha \left [ \mathcal{L}^+ \{ \hat{c}_\alpha^\dagger \hat{\rho}_{\mathrm{ss}} \hat{c}_\alpha\} \right] \hat{c}_\alpha^\dagger \right \}$, and $\delta = \mathrm{Tr} \left \{  \hat{c}_\alpha \left [ \mathcal{L}^+ \{ \hat{c}_\alpha \hat{\rho}_{\mathrm{ss}} \hat{c}_\alpha^\dagger\} \right] \hat{c}_\alpha^\dagger \right \}$. 

\begin{figure}[h!]
    \centering
    \includegraphics[width=1.0\textwidth]{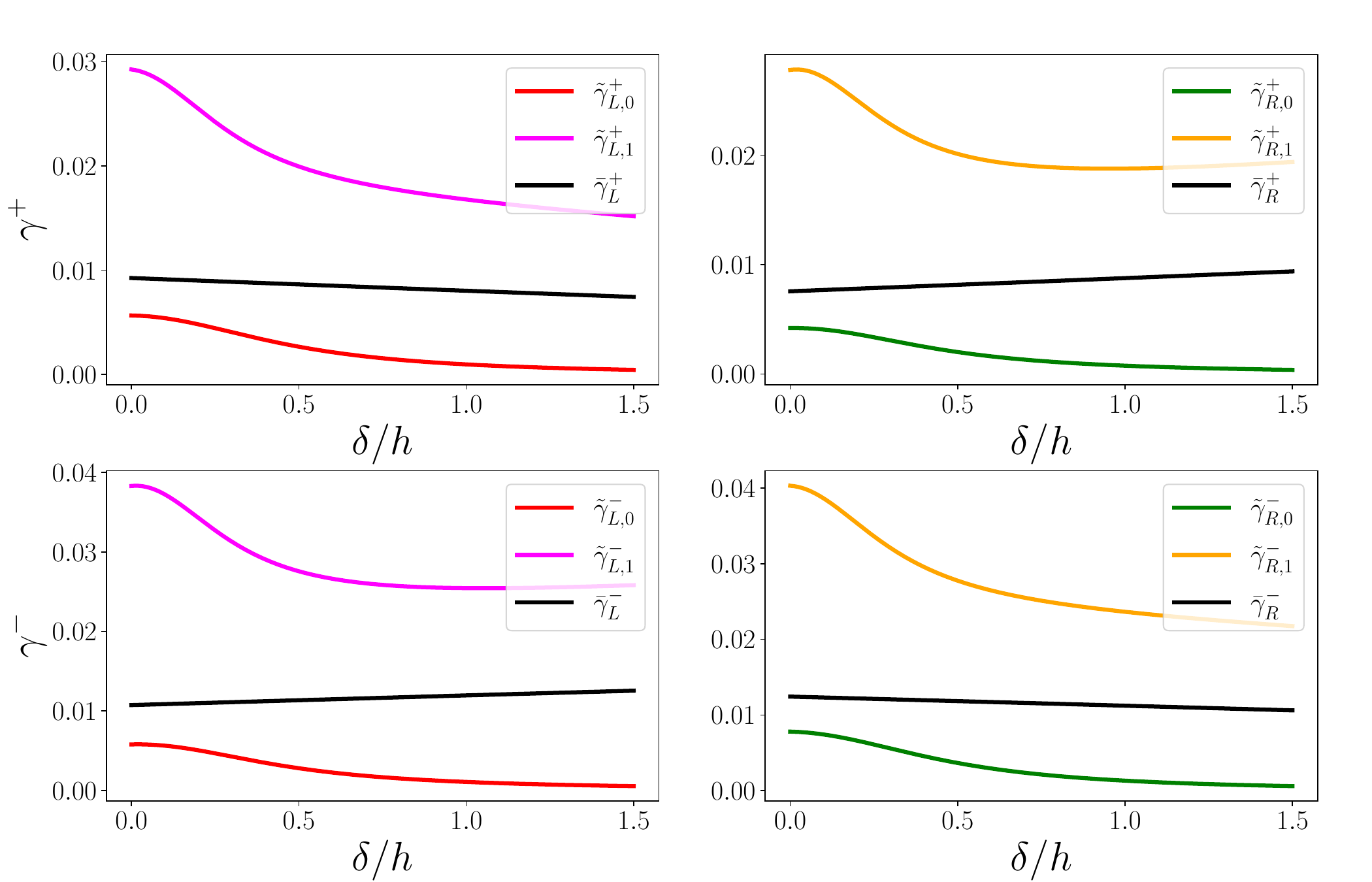}
    \caption{Decay $\gamma^-$ and gain $\gamma^+$ rates of the master equation in the site basis as a function of the local energy levels detuning  $\delta/h$. Parameters: $h=1$, $g=0.5$, $\Gamma = 0.02$, $\Omega = 100$, $T_L=T_R = 2h$, $\mu_L/h=0.7$, and  $\mu_R/h=0$. $\bar{\gamma}^{\pm}_{\alpha}$ denotes the rates of the Lindblad generators in Eq.~\eqref{eq:generatorlocal_s}, while $\tilde{\gamma}_{\alpha, 0/1}$ are the two solutions of Eq.~\eqref{eq:final}.}
    \label{fignonlocaljumps}
\end{figure}

This system of equations can have up to two real solutions $\tilde{\gamma}^\pm_{0,1}$, which is illustrated in Fig.~\ref{fignonlocaljumps}, where the computed rates are compared with the rates of the total dissipator and the quantities in the left-hand side are obtained using the Landauer-B\"uttiker approach.   
For these two solutions $\tilde{\gamma}^\pm_{0,1}$, one can construct the map
$\mathcal{M}_0^{\rm c, j}$ 
\begin{equation}
\label{Mc_noclick}
    \mathcal{M}_0^{\rm c, j} = e^{ ( \mathcal{L}_{\rm j}^{\rm c}+\mathcal{L}_{LR}^{\rm c}-i \mathcal{H}_{\text{eff}}^{\rm c})\Delta t},
\end{equation} 
where 
\begin{align}
\label{eq_jump_appendix}
    \mathcal{L}_{\rm j}^{\rm c} \hat{\rho}_S = \bar{\gamma}^{-}_{L}(1 - \eta_1^{\rm c}) \hat{c}_L \hat{\rho}_S \hat{c}_L^\dagger + \bar{\gamma}^{+}_{L}(1 -\eta_2^{\rm c}) \hat{c}_L^\dagger \hat{\rho}_S \hat{c}_L + \bar{\gamma}^{-}_{R}(1  - \eta_3^{\rm c}) \hat{c}_R \hat{\rho}_S \hat{c}_R^\dagger +\bar{\gamma}^{+}_{R}(1  - \eta_4^{\rm c}) \hat{c}_R^\dagger \hat{\rho}_S  \hat{c}_R ,
\end{align}

Now assuming that there is an unravelling with eights jump operators in the energy eigenbasis: $\hat{L}_1^1 = \sqrt{\gamma^{-}_{L,1} \eta^d_1} \hat{d}_1$, $\hat{L}_2^1 = \sqrt{\gamma^{+}_{L,1}\eta^d_2} \hat{d}^\dagger_1$, $\hat{L}_3^1 = \sqrt{\gamma^{-}_{L,2}\eta^d_3} \hat{d}_2$, $\hat{L}_4^1 = \sqrt{\gamma^{+}_{L,2}\eta^d_4} \hat{d}^\dagger_2$, $\hat{L}_5^1 = \sqrt{\gamma^{-}_{R,1}\eta^d_5} \hat{d}_1$, $\hat{L}_6^1 = \sqrt{\gamma^{+}_{R,1}\eta^d_6} \hat{d}^\dagger_1$, $\hat{L}_7^1 = \sqrt{\gamma^{-}_{R,1}\eta^d_7} \hat{d}_2$, and $\hat{L}_8^1 = \sqrt{\gamma^{+}_{R,2}\eta^d_8} \hat{d}^\dagger_2$, with $\eta^d_p = \tilde{\Gamma}^\pm_{\alpha, j}/\gamma^{\pm}_{\alpha, j}$. The steady-state current and fluctuations can be subsequently written as
\begin{eqnarray}
\begin{cases}
\langle J \rangle_\alpha = \sum_j \tilde{\Gamma}^{+}_{\alpha,j} - (\tilde{\Gamma}^{+}_{\alpha,j} + \tilde{\Gamma}^{-}_{\alpha,j}) \langle \hat{n}_j \rangle_{ss}\\
\langle \langle J^2 \rangle \rangle_\alpha = \sum_j \tilde{\Gamma}^{+}_{\alpha,j} - (\tilde{\Gamma}^{+}_{\alpha,j} - \tilde{\Gamma}^{-}_{\alpha,j}) \langle \hat{n}_j \rangle_{ss} - \zeta_{\text{d},j}~ (\tilde{\Gamma}^{+}_{\alpha, j})^2 + (\beta_{\text{d},j} + \xi_{\text{d},j})~ \tilde{\Gamma}^{+}_{\alpha, j}\tilde{\Gamma}^{-}_{\alpha, j} - \delta_{\text{d},j}(\tilde{\Gamma}^{-}_{\alpha,j})^2,   
\label{eq:final_d}    
\end{cases}
\end{eqnarray}
where $\langle \hat{n}_j \rangle_{ss} = \mathrm{Tr} \{ \hat{d}_j^\dagger\hat{d}_j \hat{\rho}_{\mathrm{ss}} \}$, $\zeta_{\text{d},j} = \mathrm{Tr} \left \{  \hat{d}_j^\dagger \left [ \mathcal{L}^+ \{ \hat{d}_j^\dagger \hat{\rho}_{\mathrm{ss}} \hat{d}_j\} \right] \hat{d}_j \right \}$, $\beta_d = \mathrm{Tr} \left \{  \hat{d}_j^\dagger \left [ \mathcal{L}^+ \{ \hat{d}_j \hat{\rho}_{\mathrm{ss}} \hat{d}_j^\dagger\} \right] \hat{d}_j \right \}$, $\xi_{\text{d},j} = \mathrm{Tr} \left \{ \hat{d}_j \left [ \mathcal{L}^+ \{ \hat{d}_j^\dagger \hat{\rho}_{\mathrm{ss}} \hat{d}_j\} \right] \hat{d}_j^\dagger \right \}$, and $\delta_{\text{d},j} = \mathrm{Tr} \left \{ \hat{d}_j \left [ \mathcal{L}^+ \{ \hat{d}_j \hat{\rho}_{\mathrm{ss}} \hat{d}_j^\dagger\} \right] \hat{d}_j^\dagger \right \}$. 

The system of equations~\eqref{eq:final_d} comprises two equations for the four variables $\tilde{\Gamma}^\pm_{\alpha, j}$, and is therefore underdetermined. The rates $\tilde{\Gamma}^\pm_{\alpha, j}$ cannot be determined uniquely from these equations alone.

One of the solutions of the system of equations \eqref{eq:final}, with the rates smaller than the rates of the master equation $\bar{\gamma}^{\pm}_{\alpha} > \tilde{\gamma}^{\pm}_{\alpha}$, results in completely positivity of  $\mathcal{M}_0^{\rm c, j}$ in some regimes but not in general, as discussed in the main text and illustrated in Fig~\ref{fig:cpness_of_imd}.

\begin{figure}
    \centering
    \includegraphics[width=0.5\textwidth]{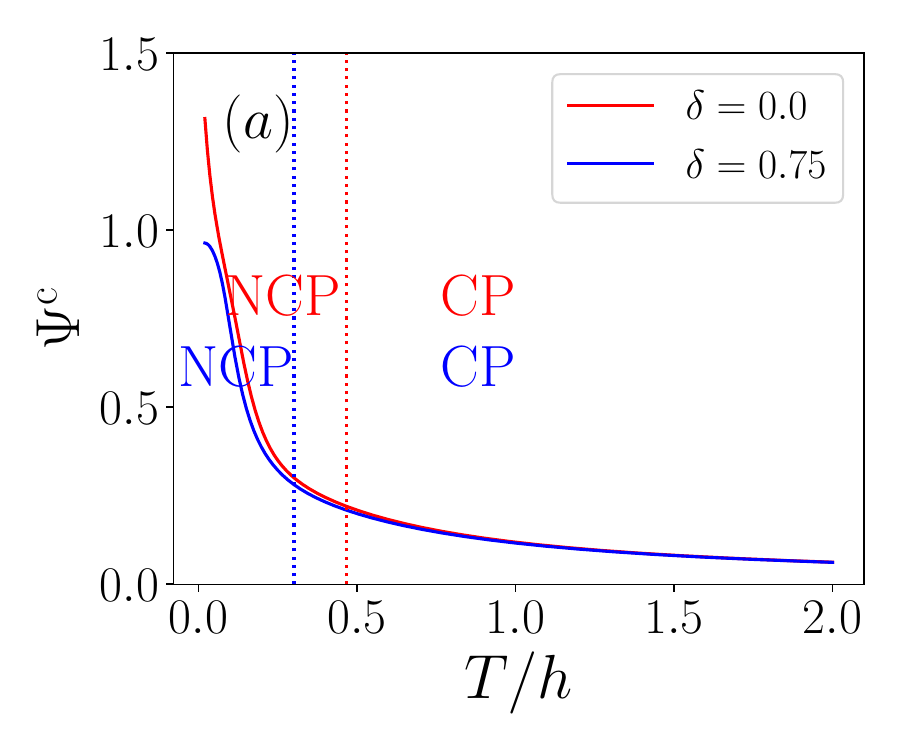}
    \caption{Average ratio of the interference and  local rates $ \Psi^{\rm c}$ in the limit $\Gamma \ll \Delta \ll \Omega$ as a function of the temperature $T/h$. Parameters: $h=1$, $g=0.5$, $\Gamma = 0.02 $, $\Omega = 100$,$\mu_L/h = 0.7$, and $\mu_R/h=0$. The vertical lines indicate the temperatures at which the map $\mathcal{M}_0^{\rm c, j}$ becomes CP. For simplicity, lines drawn in the same colour correspond to the same value of the detuning $\delta$.}
    \label{fig:approximation}
\end{figure}

Before concluding this section, we show that if the interference rates $\bar{\Lambda}_{LR}^\pm$ and $\bar{\Lambda}_{RL}^\pm$ are small compared to single-site rates $\bar{\gamma}_\alpha^\pm$, then the map $\mathcal{M}_0^{\rm c, j}$ is CP; if they are comparable with them, then $\mathcal{M}_0^{\rm c, j}$ is not CP. To this end, we introduce the quantities $\Psi^{\rm{c}} = (|\bar{\Lambda}_{LR}^+|  |\bar{\Lambda}_{LR}^-|)^{1/2}/(\bar{\gamma}_L^+\bar{\gamma}_L^- \bar{\gamma}_R^+\bar{\gamma}_L^-)^{1/4}$, which represents geometric means of the interference and single-site rates. They are plotted in Fig.~\ref{fig:approximation}.

As illustrated in Fig~\ref{fig:cpness_of_imd}, the map $\mathcal{M}_0^{\rm c, j}$ fails to be CP at small temperatures. Indeed, when $\Psi^{\rm{c}} \ll 1$, the interference term $\mathcal{L}^{c}_{LR}$ is negligible, so the map $\mathcal{M}_0^{\rm c, j}$ is CP $-$ corresponding to an additive unravelling that can be used to assess current fluctuations.   

\section{Existence of additive unravellings under non-Markovian evolution}
\label{appsec:non_markovianity}

Any local-in-time dissipator $\mathcal{L}$ can be transformed to a canonical Lindblad-like form~\cite{Hall_2014}. Indeed, let us diagonalise the rate matrices $\Lambda^{\pm}$ by a unitary rotation $[\mathrm{U}^{\pm}]^\dagger\Lambda^{\pm}\mathrm{U}^{\pm}=\mathrm{diag}[\lambda_1^{\pm},\lambda_2^{\pm}]$, so the dissipator $\mathcal{L}$ acquires its canonical form
\begin{equation}
\mathcal{L} = \sum_{j=1}^2\sum_{s= \pm} \lambda^s_j \mathcal{D}[\hat{L}^s_j],
\label{eq:canonical_form}
\end{equation}
where
$\mathcal{D}[\hat{L}^s_j]$ is a Lindblad dissipator, $\lambda^s_j$ are the eigenvalues of the rate matrix $\Lambda^{s}$, and $\hat{L}^s_j$ are the jump operators that are defined as follows $\hat{L}^-_j= \sum_k U_{jk}^{-} \hat{d}_k$ and $\hat{L}^+_j= \sum_k [U_{jk}^{+}]^* \hat{d}^\dagger_k$. These jump operators correspond to the loss and gain of excitations relative to modes set by the eigenbases of the matrices $\Lambda^{\pm}$. In general, the loss and gain jump modes are different from each other, except when $[\Lambda^+,\Lambda^-]=0$, in which case there exist gain and loss operators such that $\hat{L}^+_j = [\hat{L}^-_j]^\dagger$. Importantly, the jump operators $\hat{L}^\pm_j$ make up an orthonormal basis set of traceless operators, namely,
\begin{equation}
 \operatorname{Tr}\{ \hat{L}^s_j\} =0, \quad \operatorname{Tr}\{ \hat{L}^s_j \hat{L}^{r\dagger}_k \} =\delta_{jk} \delta_{sr}.  
\label{eq:canonical_operators}
\end{equation}

However, a dissipator $\mathcal{L}$ not always can be cast in Lindblad form. In particular, the overall dissipator $\mathcal{L}$ for our setup may not be reduced to Lindblad form for certain values of the parameters since $\mathcal{L}^{\rm m}_{LR}$ can never be represented in Lindblad form themselves. Owing to this, one or more negative eigenvalues $\lambda_j^\pm$ may appear in the canonical representation of $\mathcal{L}$~(Eq.\eqref{eq:canonical_form}), which, in turn, indicates the onset of non-Markovian evolution, as was shown by Rivas et al.~\cite{Rivas_2010}. Following this, one can characterise the extent of asymptotic non-Markovianity (ANM) by~\cite{Mitchison_2018, Hall_2014}   
\begin{equation}
\nu_{\infty} = \sum_{s=\pm} \sum_{j=1,2} \max[0, -\lambda_{j, \infty}^s].
\label{asymptotic_non_markovianity}
\end{equation}

Both $\mathcal{L}^{\rm c}_\alpha$ and $\mathcal{L}^{\rm d}_\alpha$ are in Lindblad form, so that presence of ANM is related to the cross-terms $\mathcal{L}^{\rm m}_{LR}$. As a result, ANM captures the violation of additivity in the local and global bases.

\begin{figure}[!t]
    \centering
    \includegraphics[width=0.5\textwidth]{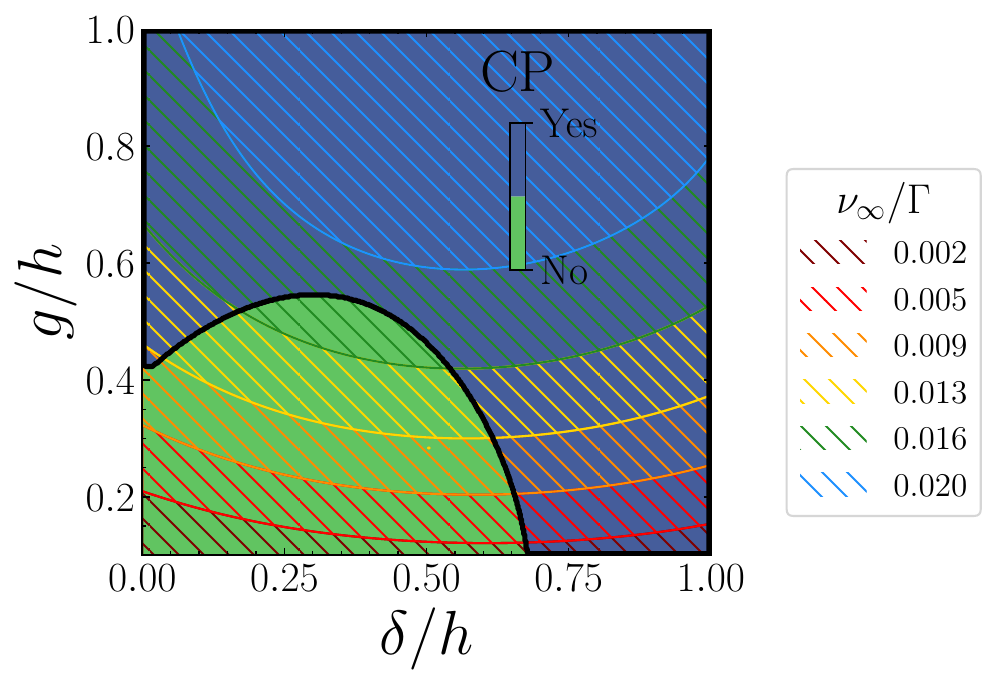}
    \caption{Complete positivity (CP) of the quantum map $\mathcal{M}_0^{\rm c,j}$ in the limit $\Gamma \ll \Delta \ll \Omega$ as a function of the local energy levels detuning $\delta/h$ and interdot coupling $g/h$ in the presence of asymptotic non-Markovianity. Parameters: $h=1$, $\Gamma = 0.02$, $\Omega = 100$, $\mu_L/h = 0.7$, $\mu_R=0$, $T = T_L=T_R = 0.01$. The blue colour denotes that the quantum map is CP and the obtained rates $\tilde{\gamma}^{\pm}_{\alpha}$ are smaller than the actual rates $\bar{\gamma}^{\pm}_{\alpha}$, i.e. $\bar{\gamma}^{\pm}_{\alpha} > \tilde{\gamma}^{\pm}_{\alpha}$ (``Yes"), while the green one corresponds to a non-CP quantum map (``No") with positive or negative obtained rates. This plot is invariant with respect to $\Delta t$, which we verified for $1001$ values of $\Delta t \in [10^{-3},10^3]$.}
    \label{fig:anm_vs_coupling_detuning}
\end{figure}

Asymptotic non-Markovianity can exist alongside with additive unravellings of non-additive master equation~\eqref{eq:exact_master_equation} that can be used to assess current fluctuations in the case of imperfect detection, which is illustrated in Fig.~\ref{fig:anm_vs_coupling_detuning}.

As illustrated in Fig.~\ref{fig:anm_vs_coupling_detuning}, steady-state current fluctuations can be computed by assuming additive unravellings of the non-additive master equation~\eqref{eq:exact_master_equation} even in the presence of asymptotic non-Markovianity. This Markovian unravelling based on imperfect detection should be distinguished from Ref.~\cite{Piilo_2008} that constructs an explicitly non-Markovian quantum-trajectory framework for general time-local QMEs.

\end{document}